\documentclass[]{emulateapj}
\usepackage{natbib}
\usepackage{apjfonts}
\usepackage{epsfig}

\newcommand{\gas}{\mathrm{gas}}
\newcommand{\dyn}{\mathrm{dyn}}
\newcommand{\LCDM}{\Lambda\mathrm{CDM}}
\newcommand{\OmegaL}{\Omega_{\Lambda}}
\newcommand{\OmegaM}{\Omega_{\mathrm{M}}}
\newcommand{\Omegab}{\Omega_{\mathrm{b}}}

\newcommand{\Mvir}{M_{\mathrm{vir}}}
\newcommand{\Vvir}{V_{\mathrm{vir}}}
\newcommand{\Rvir}{R_{\mathrm{vir}}}
\newcommand{\Cvir}{C_{\mathrm{vir}}}

\newcommand{\Mg}{\mathrm{Mg}_{2}}
\newcommand{\tdisk}{t_{\mathrm{disk}}}

\newcommand{\Mdyn}{M_{\dyn}}
\newcommand{\Mstar}{M_{\star}}
\newcommand{\Mgas}{M_{\gas}}
\newcommand{\Msun}{M_{\sun}}
\newcommand{\Mtotal}{M_{\mathrm{total}}}
\newcommand{\re}{R_{\mathrm{e}}}

\newcommand{\sigmae}{\sigma}
\newcommand{\Ie}{I_{\mathrm{e}}}

\newcommand{\sigmaRV}{\sigma_{\mathrm{RV}}}
\newcommand{\sigmaIV}{\sigma_{\mathrm{IV}}}
\newcommand{\sigmaIR}{\sigma_{\mathrm{IR}}}
\newcommand{\sigmaRR}{\sigma_{\mathrm{RR}}}
\newcommand{\sigmaVV}{\sigma_{\mathrm{VV}}}
\newcommand{\sigmaII}{\sigma_{\mathrm{II}}}
\newcommand{\sigmaXY}{\sigma_{\mathrm{XY}}}
\newcommand{\scatter}{\left<\Delta^{2}\right>^{1/2}}

\newcommand{\DM}{\mathrm{DM}}
\newcommand{\MBH}{M_{\mathrm{BH}}}
\newcommand{\Mdm}{M_{\DM}}
\newcommand{\Mbulge}{M_{\mathrm{bulge}}}
\newcommand{\Mdisk}{M_{\mathrm{disk}}}
\newcommand{\mb}{m_{\mathrm{b}}}

\newcommand{\msigma}{\MBH\mathrm{-}\sigmae}

\newcommand{\etatherm}{\eta_{\mathrm{therm}}}

\newcommand{\remstar}{\re-\Mstar}

\newcommand{\cdf}{\bar{f}}
\newcommand{\ccdf}{s(\cdf)}
\newcommand{\fgas}{f_{\gas}}
\newcommand{\fstar}{f_{\star}}
\newcommand{\fmtl}{\Mtotal/\Mstar}
\newcommand{\mtl}{M/L}
\newcommand{\qEOS}{q_{\mathrm{EOS}}}

\newcommand{\rd}{r_{\mathrm{d}}}
\newcommand{\rperi}{r_{\mathrm{peri}}}

\shorttitle{Gas Dissipation \& the FP}
\shortauthors{Robertson et al.}

\begin{document}

\title{The Fundamental Scaling Relations of Elliptical Galaxies}
\author{Brant Robertson\altaffilmark{1,5},
	Thomas J. Cox\altaffilmark{1},
	Lars Hernquist\altaffilmark{1},
	Marijn Franx\altaffilmark{2},\\
	Philip F. Hopkins\altaffilmark{1},
        Paul Martini\altaffilmark{3},
        Volker Springel\altaffilmark{4}}

\altaffiltext{1}{Harvard-Smithsonian Center for Astrophysics, 
        60 Garden St., Cambridge, MA 02138, USA.}
\altaffiltext{2}{Leiden Observatory, P.O. Box 9513, NL-2300 RA Leiden, 
        Netherlands.}
\altaffiltext{3}{The Ohio State University, Department of Astronomy,
        140 West 18th Ave., Columbus, OH 43210, USA.}
\altaffiltext{4}{Max-Planck-Institut f\"ur Astrophysik, Karl-Schwarzschild-Stra\ss e  1, 
	85740 Garching bei M\"unchen, Germany.}
\altaffiltext{5}{brobertson@cfa.harvard.edu}

\begin{abstract}

We examine the fundamental scaling relations of elliptical galaxies
formed through mergers. Using hundreds of simulations to judge the
impact of progenitor galaxy properties on the properties of merger
remnants, we find that gas dissipation provides an important
contribution to tilt in the Fundamental Plane relation.
Dissipationless mergers of disks produce remnants that occupy a plane
similar to that delineated by the virial relation.  As the gas content
of progenitor disk galaxies is increased, the tilt of the resulting
Fundamental Plane relation increases and the slope of the $\remstar$
relation steepens.  For gas fractions $\fgas > 30\%$, the simulated
Fundamental Plane scalings ($\re \propto \sigma^{1.55} \Ie^{-0.82}$)
 approach those observed in the $K$-band ($\re \propto \sigma^{1.53}\Ie^{-0.79}$).
The dissipationless
merging of spheroidal galaxies and the re-merging of disk galaxy
remnants roughly maintain the tilt of the Fundamental Plane occupied
by the progenitor ellipticals, approximately independent of the
orbital energy or angular momentum.  Dry merging of spheroidal systems
at redshifts $z<1$ is then expected to maintain the stellar-mass
Fundamental Plane relations imprinted by gas-rich merging during the
epoch of rapid spheroid and supermassive black hole growth at
redshifts $z \approx 1-3$.  In our simulations, feedback from
supermassive black hole growth has only a minor influence on the
stellar-mass scaling relations of spheroidal galaxies, but may play a
role in maintaining the observed Fundamental Plane tilt at optical
wavelengths by suppressing residual star formation in merger remnants.

We estimate that $\approx 40-100\%$ of the Fundamental Plane tilt induced by 
structural properties, as opposed to stellar population effects, 
owes to trends
in the central total-to-stellar mass ratio $\fmtl$ produced by
dissipation.  Gas cooling allows for an increase in central stellar
phase-space density relative to dissipationless mergers, thereby
decreasing the central $\fmtl$.  Lower mass systems obtain greater
phase-space densities than higher mass systems, producing a galaxy
mass-dependent central $\fmtl$ and a corresponding tilt in the
Fundamental Plane.  We account for these trends in the importance of
dissipation with galaxy mass in terms of the inefficient cooling of
collisionally heated gas in massive halos and dynamically varying gas
consumption timescales in smaller systems.
\end{abstract}


\section{Introduction}
\label{section:introduction}

Elliptical galaxies represent a fascinating combination of complexity
and regularity.  A leading theory for the origin of early-type
galaxies is based on mergers of disk galaxies \citep{toomre1972a,
toomre1977a} and likely involves gas
dissipation, star formation, and supermassive black hole feedback
\citep{barnes1992a,barnes1992b,barnes1996a,mihos1994a,mihos1996a,
di_matteo2005a} in addition to stellar dynamics.  Despite
their complex origins, early-type galaxies obey a regular set of
scaling relations that connect their photometric and kinematic
properties, most notably the relation between effective radius $\re$,
central stellar velocity dispersion $\sigmae$, and average central
surface brightness $\Ie$ known as the Fundamental Plane 
\citep[FP;][]{dressler1987a,djorgovski1987a}

\begin{equation}
\label{eqn:fp}
\re \propto \sigmae^{\alpha} \Ie^{-\beta}.
\end{equation}
\noindent
The virial theorem can be used to calculate this relation for homologous 
systems, which gives $\alpha=2$, $\beta=1$ \citep[e.g.,][hereafter the 
``virial scaling'']{faber1987a}.

While the observational determination of the FP was originally
motivated as a precise distance indicator to improve upon the
previously known luminosity ($L$) -- velocity dispersion ($\sigmae$)
relation \citep[][]{ faber1976a}, the importance of the FP
scalings and its correspondingly small scatter for theories of
elliptical galaxy formation was also realized.  The first
observationally determined FP scalings 
\citep[$\alpha \sim 1.3- 1.4$,$\beta \sim 0.8-0.9$, at optical wavelengths;][]{dressler1987a,djorgovski1987a}
differed from the virial scaling, indicating a
``tilt'' relative to the expectation for homologous systems.  The FP
tilt implied that the mass--to--light ratio $M/L$ likely varies as a
function of galaxy mass or luminosity as
\begin{equation}
\label{eqn:mtl}
\frac{M}{L} \propto L^{\gamma},
\end{equation}
\noindent
within the range $\gamma \approx 1/5 - 1/4$.  \cite{faber1987a} noted
that deviations from the FP can be induced by $\mtl$ variance owing to
e.g. metallicity or age trends in stellar populations, dynamical or
structural properties, and the relative distribution of dark and
baryonic matter.  In principle, each of these effects may also
introduce a systematic tilt into the observed FP if they vary as a
function of elliptical galaxy mass.

The purpose of the current paper is to gauge the importance of various
contributions to the tilt in the observed FP in the context of the
scenario where elliptical galaxies form from mergers. Using hundreds
of simulated galaxy mergers that include the physics of gas cooling,
star formation, supernova feedback, and black hole accretion and
feedback, we determine that gas dissipation may significantly
contribute to the tilt of the observed FP, in addition to tilt induced
by $\mtl$ trends from stellar populations.  We propose that elliptical
galaxies initially form in gas-rich mergers from disk galaxy
progenitors whose gas fractions exceed $\fgas \sim 30\%$, and show
that these remnants display substantial FP tilt.

We connect the origin of this tilt to the central stellar phase-space
density of the remnants.  In small mass systems where dissipation is
most important, the stellar phase-space density of remnants increases
substantially in high-gas fraction mergers.  The central stellar
phase-space density in massive ellipticals remains similar in mergers
with varying gas fractions, with their stars on average obtaining
lower phase-space densities than reached in smaller systems.  This
mass-dependent phase-space density trend translates into a
mass-dependent trend in the ratio of total mass to stellar mass
$\fmtl$ in the central regions of ellipticals and a corresponding tilt
in the FP.  We then explain this mass-dependent importance of
dissipation in terms of the inefficient cooling of collisionally
heated gas in massive ellipticals and the dynamically varying gas
consumption timescale of smaller systems.  It is interesting, and
possibly significant, that the gas fraction required to reproduce the
observed tilt in the FP is, as discussed by \cite{hernquist1993a}
similar to that needed for mergers of disks to yield
the high central phase space densities of ellipticals.

In accord with other work
\citep[e.g.,][]{capelato1995a,dantas2003a,nipoti2003a,boylan-kolchin2005a},
we find that subsequent dissipationless merging between spheroidal
galaxies roughly maintains the FP tilt.  Moreover, we find that a
single generation of dissipationless re-merging of remnants will
induce scatter in the $\msigma$ relation but will not destroy the
correlation, as spheroidal galaxy mergers also do not dramatically
alter the $\re-\Mstar$ relation when viewed as a mass-sequence.
Possible dry merging between spheroidal galaxies at redshifts $z<1$ as
indicated by recent observations \citep{bell2005a,van_dokkum2005a} is
thus not expected to destroy tight FP or $\msigma$ relations generated
during spheroid formation through gas-rich mergers at higher redshifts.

This paper is organized as follows.  We review the observational and
theoretical work on elliptical galaxy formation and scaling relations
in \S \ref{section:review}.  We present our methodology in \S
\ref{section:methodology} and our results in \S \ref{section:results}.
We discuss the implications of our work in \S \ref{section:discussion}
and summarize and conclude in \S \ref{section:summary}.  Throughout,
we adopt a flat $\LCDM$ universe with $\OmegaM=0.3$, $\OmegaL=0.7$,
$\Omegab=0.04$, and a Hubble parameter $H_{0} = 100h$ km s$^{-1}$
Mpc$^{-1}$ with $h=0.7$.

\section{Review of Elliptical Galaxy Scaling Laws}
\label{section:review}

Scaling laws describing the regularity of the properties of elliptical
galaxies have been known since at least \cite{fish1964a}, who reported
a relation between their potential energy $W$ and mass $\Mstar$ as $W
\propto \Mstar^{3/2}$.  Even earlier, \citet[][hereafter dV]{de_vaucouleurs1948a}
had demonstrated that ellipticals generally
have a surface brightness profile $\log I(r) \propto r^{1/4}$, at
least over some range in radius $r$, and implications of the dV
profile for the mass-dependent properties of ellipticals were
recognized before the \cite{fish1964a} paper
\citep[e.g.,][]{poveda1958a}.

\cite{sandage1972a} found a color--magnitude relation for Virgo and
Coma cluster ellipticals \citep[see also][]{stebbins1952a,de_vaucouleurs1961a}.
\cite{faber1973a}
discovered a similar color--magnitude trend in Local Group and cluster
ellipticals, as well as an absorption-line strength--magnitude
relation, and suggested that these properties of elliptical galaxies
depend primarily on galaxy luminosity.  \cite{faber1976a} established
the relation between luminosity $L$ and velocity dispersion $\sigma$,
providing further evidence that ellipticals follow a regular sequence
as a function of mass.  \cite{kormendy1977a} showed that the surface
brightnesses and effective radii of ellipticals correlate with galaxy
luminosity and with one another \citep[see also][]{binggeli1984a}.

Important early indications that a second parameter (in addition to
mass) governs the properties of ellipticals came with the
\cite{terlevich1981a} and \cite{tonry1981a} work that implied a
correlation of $L-\sigma$ and absorption-line strength-- luminosity
relation ($\Mg-L$) residuals.  While these findings were later
contradicted \citep[e.g.][]{dressler1984a}, that elliptical galaxies
were not a one-parameter family remained an important possibility.

The discovery of the FP \citep{dressler1987a,djorgovski1987a}
definitively revealed that elliptical galaxy properties are set by at
least two parameters.  Specifically, ellipticals were found to obey a
relation between $\re$, $\sigma$, and $\Ie$ as given by Equation
(\ref{eqn:fp}), with less than half the scatter of the
\cite{faber1976a} $L-\sigma$ relation.  The small
scatter of the FP was immediately noticed, implying that the process of
elliptical galaxy formation 
must result in a very
regular mass-sequence.  \cite{faber1987a} noted that mass-to-light
ratio ($M/L$) variations can influence observations of the FP by
inducing tilt relative to the plane defined by the virial relation.
Much of the subsequent work on the FP has centered around possible
causes of $M/L$ variation or other origins for tilting the FP relative
to the virial plane.

Numerous subsequent observations verified and improved the FP relation for ellipticals
\citep{lucey1991a,lucey1991b,de_carvalho1992a,bender1992a,jorgensen1992a,guzman1993a,jorgensen1993a,saglia1993a,bender1994a,prugniel1994a,pahre1995a,jorgensen1996a,prugniel1996a,busarello1997a,graham1997a,bender1998a,pahre1998a,pahre1998b,mobasher1999a,kronawitter2000a,gerhard2001a,bernardi2003c,padmanabhan2004a,woo2004a,cappellari2005a} and extended the FP
determinations to higher redshifts \citep{franx1993a,van_dokkum1996a,kelson1997a,schade1997a,van_dokkum1998a,jorgensen1999a,treu1999a,kelson2000a,kelson2000b,kelson2000c,kochanek2000a,treu2001a,van_dokkum2001a,van_dokkum2001b,treu2002a,gebhardt2003a,rusin2003a,van_dokkum2003a,van_dokkum2003b,van_der_wel2004a}.  While the specific details may vary, these works generally conclude that
\begin{itemize}
\item A tight fundamental plane between the elliptical galaxy properties $\re$, $\sigma$, $\Ie$ 
exists, and extends in some form to at least redshift $z\sim1$.
\item Ellipticals are old, with formation redshifts $z>1$, and their color 
evolution, which controls the FP normalization, is roughly
consistent with passive evolution of their stellar populations.
\item Some of the FP tilt must originate from the change in stellar population $M/L$ with galaxy
mass, but the extent to which stellar populations contribute to $M/L$-induced FP tilt is
debated.
\end{itemize}
Of specific interest to modelers are the observations that made
definite statements about the influence of structural or kinematic
nonhomology on $M/L$ and the FP tilt, especially those that conclude
directly that these nonhomologies are either unimportant or of minor 
significance
\citep{gerhard2001a,cappellari2005a}, significant
\citep[e.g.][]{padmanabhan2004a}, or of possible but as yet not fully
determined importance \citep[e.g.][]{pahre1998a}.

Moreover, elliptical galaxy properties related to the FP or its
projections have been extensively observed.  These include studies of
the photometric profiles
\citep[e.g.][]{de_vaucouleurs1948a,sersic1968a,kormendy1982a,burkert1993a,caon1993a},
internal kinematic structure
\citep[e.g.][]{binney1978a,davies1983a,davies1988a,bender1994a},
metallicity
\citep{de_carvalho1992a,bender1993a,bender1996a,bernardi1998a,bernardi2003d},
and the $\msigma$ relation
\citep[e.g.][]{gebhardt2000a,ferrarese2000a,tremaine2002a}. An
important, related property is the size-stellar mass relation
\citep{shen2003a,trujillo2004a,trujillo2004b,mcintosh2005a}, that
indicates a power-law correlation between some characteristic galaxy
size and the stellar mass or luminosity.

Interpreting these observations has been the focus of various
theoretical efforts \citep[see, e.g.][for a description of early
results]{barnes1992b}.  Notably, simulations of the formation of
ellipticals and their properties in the context of the merger
hypothesis have been performed using a variety of progenitor models
including spheroidal
\citep{white1979a,capelato1995a,dantas2003a,gonzalez-garcia2003a,nipoti2003a},
disk
\citep{toomre1972a,gerhard1981a,farouki1982a,farouki1983a,barnes1991a,barnes1992a,hernquist1992a,hernquist1993a,hernquist1993b,mihos1994a,mihos1996a,hibbard1996a,bekki1998a,dubinski1998a,naab1999a,naab2001a,aceves2005a},
and cosmological systems \citep[e.g.][]{aarseth1980a,saiz2004a}.
Analytical models of elliptical galaxies have also been formulated
\citep{hernquist1990a,ciotti1991a,ciotti1996a,ciotti1996b,ciotti1999a},
aiding the interpretation of both the observational and simulation
results.  In what follows, we combine features of many of these
previous theoretical endeavors by simulating mergers between
dissipational and dissipationless disk galaxies, spheroidal systems,
and merger remnants to determine their fundamental scaling relations.

\section{Methodology}
\label{section:methodology}

Galaxy merging combines much complex physics, including e.g. the
collisionless dynamics of dark matter and stars, gas dissipation, star
formation, and feedback from supernovae and black hole growth.  While
each of these processes may in principle be essential for determining
merger remnant properties, their relative importance has not been
fully established.  By comparing the structure of merger remnants in
simulations which systematically include or exclude various processes,
we attempt to both test the merger hypothesis and identify the most
important physical mechanisms.

To this end, we perform a set of hundreds of simulations of galaxy
mergers.  Our suite consists of three categories of disk--disk mergers
and one class of spheroid-spheroid mergers.  The first category of
simulations, referred to as ``dissipationless,'' includes only
simulations of disks consisting of collisionless stars and dark
matter.  The second category, termed ``dissipational'' mergers,
consists of simulations that account for gas cooling, star formation,
the physics of the interstellar medium, and supernova feedback.  We
refer to the third category of calculations as ``full-model''
simulations which, in addition to the processes accounted for in the
``dissipational'' mergers, also include supermassive black hole growth
and feedback.  All three categories of disk--disk mergers are derived
from a standard set of galaxy models to enable a direct comparison of
physical processes.

To perform our numerical simulations, we utilize the GADGET2 code
\citep{springel2005b}.  GADGET2 uses the smoothed particle
hydrodynamics (SPH) formalism \citep{lucy1977a,gingold1977a} in its
entropy-conserving formulation \citep{springel2002a} to calculate the
dynamical evolution of the gas and a tree-based method to compute
gravitational forces between particles \citep{barnes1986a}.

All progenitor galaxies were created using the methods described in
\cite{springel2005a}, which allow for the generation of stable
equilibrium galaxy models.  Each galaxy contains an extended dark
matter halo, and may also consist of a stellar disk, gaseous disk,
stellar bulge, and a supermassive black hole particle.  The
collisionless components of the galaxy models are required to satisfy
the Jeans equations while the structure of the gas component is
determined by the equation of hydrostatic equilibrium and an integral
constraint on the surface mass density.

\begin{deluxetable*}{lcccccc} 
\tabletypesize{\scriptsize}
\tablecolumns{7} 
\tablewidth{0pc} 
\tablecaption{\label{table:models}Galaxy Mergers} 
\tablehead{ 
\colhead{Model} & \colhead{Progenitor} & \colhead{Redshift} & 
\colhead{Gas Fraction} & \colhead{ISM Pressurization} & 
\colhead{Pericentric Separation} & 
\colhead{\# of Simulations}\\
\colhead{} & \colhead{$\Vvir$ $[$km s$^{-1}$$]$} & \colhead{$z$} &
\colhead{$\fgas$} & \colhead{$\qEOS$} & \colhead{$\rperi$} &
\colhead{} }
\startdata 
\cutinhead{``Full-Model'' Simulations with Black Holes}
Local & 80, 115, 160, 226, 320, 500 & 0 & 0.4, 0.8 & 0.25, 1.0 & 2$\rd$ & 24\\
Intermediate$-z$& 80, 115, 160, 226, 320, 500 & 2,3 & 0.4, 0.8 & 0.25, 1.0 & 2$\rd$ & 48\\
High$-z$& 115, 160, 226, 320, 500 & 6 & 0.4, 0.8 & 0.25, 1.0 & 2$\rd$ & 20\\
Halo Concentrations & 160 & 0 & 0.4 & 1.0 & 2$\rd$ & 5\\
Disk Orientation& 160 & 0 & 0.4 & 1.0 &  Table \ref{table:orbits}& 14\\
Orbital Configuration& 160 & 0 & 0.4 & 1.0 & Table \ref{table:orbits}& 18\\
\cutinhead{Dissipational Simulations}
Local & 80, 115, 160, 226, 320, 500 & 0 & 0.4, 0.8 & 0.25, 1.0 & 2$\rd$ & 24\\
Intermediate$-z$& 80, 115, 160, 226, 320, 500 & 2,3 & 0.4, 0.8 & 0.25, 1.0 & 2$\rd$ & 48\\
High$-z$& 115, 160, 226, 320, 500 & 6 & 0.4, 0.8 & 0.25, 1.0 & 2$\rd$ & 20\\
$\fgas$ Runs & 80, 115, 160, 226, 320, 500 & 0 & 0.01, 0.025, 0.05 & 0.25, 1.0 & 2$\rd$ & 72\\
& & & 0.1, 0.2, 0.4 & & & \\
\cutinhead{Dissipationless Simulations}
Local & 80, 115, 160, 226, 320, 500 & 0 & 0.0 & -- & 2$\rd$ & 6\\
Intermediate$-z$& 80, 115, 160, 226, 320, 500 & 2,3 & 0.0 & -- & 2$\rd$ & 12\\
High$-z$& 80, 115, 160, 226, 320, 500 & 6 & 0.0 & -- & 2$\rd$ & 6\\
Wide Orbit& 80, 115, 160, 226, 320, 500 & 0 & 0.0 & -- & $0.4\Rvir$ & 6\\
Wide Orbit, Int.$-z$& 80, 115, 160, 226, 320, 500 & 2,3 & 0.0 & -- & $0.4\Rvir$ & 12\\
Wide Orbit, High$-z$& 80, 115, 160, 226, 320, 500 & 6 & 0.0 & -- & $0.4\Rvir$ & 6\\
Bulge, Local & 80, 115, 160, 226, 320, 500 & 0 & 0.0 & -- & 2$\rd$ & 6\\
Bulge, Intermediate$-z$& 80, 115, 160, 226, 320, 500 & 2,3 & 0.0 & --  & 2$\rd$ & 12\\
Bulge, High$-z$& 80, 115, 160, 226, 320, 500 & 6 & 0.0 & -- & 2$\rd$ & 6\\
High-Res & 80, 115, 160, 226, 320, 500 & 0 & 0.0 & -- & 2$\rd$ & 6\\
\cutinhead{Spheroid Simulations}
Local & 80, 115, 160, 226, 320, 500 & 0 & 0.0 & -- & $0.025\Rvir$ & 6\\
Wide/Elliptical Orbit& 80, 115, 160, 226, 320, 500 & 0 & 0.0 & -- & $0.4\Rvir$ & 6\\
Full-Model Re-mergers & 80, 115, 160, 226 & 0 & 0.4 & 0.25 & $0.05\Rvir$ & 4
\enddata 
\end{deluxetable*}

\subsection{Dissipationless Disk Simulations}
\label{subsection:methodology:dissipationless}

The dissipationless disk progenitors consist of an exponential stellar
disk embedded in a dark matter halo with virial velocities in the
range $\Vvir = 80-500$ km s$^{-1}$.  We initialize the size of the
disk according to the \cite{mo1998a} formalism for dissipational disk
galaxy formation \citep[see also][]{fall1980a,blumenthal1986a}
assuming the disk contains a fraction of the total
galaxy angular momentum equal to its mass fraction, which we set to
$m_{\mathrm{d}} \equiv \Mdisk / \Mvir = 0.041$ to match the Milky
Way-like model used by \cite{springel2005a}.  The disk scalelength
$\rd$ is then determined by the galaxy spin $\lambda$ and the dark
matter halo concentration $\Cvir$.  We adopt $\lambda=0.033$, which is
near the mode of the redshift- and mass- independent distribution of
dark matter halo spins measured in cosmological N-body simulations
\citep{vitvitska2002a}.  For the Navarro-Frenk-White halo
concentration $\Cvir$ \citep[][NFW]{navarro1997a}, we adopt the mass-
and redshift-dependent dark matter halo concentrations measured by
\cite{bullock2001a}
\begin{equation}
\label{eqn:cvir}
\Cvir(\Mvir,z) \approx 9 \left(\frac{\Mvir}{M_{\mathrm{coll},0}}\right)^{-0.13} 
                       \left(1+z\right)^{-1},
\end{equation}
\noindent
where $M_{\mathrm{coll},0} \sim 8 \times 10^{12} h^{-1} \Msun$ is the
linear collapse mass at redshift $z=0$.  In all cases, the stellar
disk scaleheight $h_{\mathrm{d}}=0.2\rd$, similar to the Milky Way
\citep[c.f.][]{siegel2002a}.  We model the dark matter halo with a
\cite{hernquist1990a} density profile of the form
\begin{equation}
\label{eqn:hernquist}
\rho_{\mathrm{h}}(r) = \frac{\Mdm}{2\pi}\frac{a}{r\left(r+a\right)^{3}},
\end{equation}
\noindent
where the scalelength $a(\Cvir)$ maps the \cite{hernquist1990a}
profile parameters to the appropriate NFW halo parameters 
\citep[for details, see][]{springel2005a}.

Following \cite{robertson2005b}, we scale the progenitor galaxy
properties to approximate the structure of disk galaxies appropriate
for redshifts $z=0$, $2$, $3$, and $6$.  Varying the progenitor
galaxies in this manner enables us to determine the impact of
redshift-dependent galaxy properties on the scaling laws of
ellipticals.  Keeping the virial velocity $\Vvir$ fixed with redshift,
we scale the progenitor virial mass and virial radius using the relations
\begin{equation}
\label{eqn:mvir}
\Mvir = \frac{\Vvir^{3}}{10 G H(z)}
\end{equation}
\begin{equation}
\label{eqn:rvir}
\Rvir = \frac{\Vvir}{10 H(z)} ,
\end{equation}
\noindent
where $H(z)$ is the Hubble parameter.  To suitably resolve the forces
between particles in models of higher-redshift systems, we reduce the
gravitational smoothing by $(1+z)^{-1}$.  The halo concentrations also
vary with redshift and mass according to Equation (\ref{eqn:cvir}) and
the disk scalelengths decrease with redshift through their dependence
on $\Rvir$ and $\Cvir$.  We note that the redshift-dependence of disk
scalelengths agrees well with the distribution of disk scalelengths
seen out to redshift $z \approx 1$
\citep[e.g.][]{ravindranath2004a,barden2005a}.

At each redshift, we consider three separate types of dissipationless
disk mergers.  First, we examine equal mass mergers of pure disk
galaxies on prograde-prograde coplanar parabolic orbits with the
pericentric passage distance set to $\rperi = 2\rd$.  The galaxies
each contain $60,000$ dark matter and $80,000$ stellar disk particles.
Second, we repeat each merger with a wider parabolic orbit increased
to $\rperi = 0.4\Rvir$ to judge the effect of increased orbital
angular momentum on the dissipationless merger remnants.  In addition,
we repeat each nearly-radial merger with bulge components included in
the galaxies with mass fraction $\mb \equiv \Mbulge/\Mvir = 0.1367$ to
match the Milky Way-like model used in \cite{springel2005a}.  We model
the bulges with a \cite{hernquist1990a} density profile form (see
Equation \ref{eqn:hernquist}) where we set the bulge scalelength $b =
0.2\rd$.  Each bulge contains $20,000$ particles, with the number of
disk particles reduced to $60,000$ to maintain the same mass
resolution.  Finally, we re-run all the pure disk simulations at
redshift $z=0$ with higher resolution dissipationless models with
$180,000$ dark matter particles and $120,000$ disk particles to
examine issues related to numerical resolution.  While we discuss
these tests in more detail in \S \ref{section:results}, we note here
that the large set of dissipationless simulations produces results
very consistent with the restricted set of higher resolution runs.  In
all we perform $78$ dissipationless simulations, and we provide a complete
listing in Table \ref{table:models}.

\begin{deluxetable}{cccccc}
\tabletypesize{\scriptsize}
\tablecolumns{6} 
\tablewidth{0pt}
\tablecaption{\label{table:orbits}Orbital Variations
}
\tablehead{
\colhead{Models} & \colhead{$\theta_{1}$} & \colhead{$\phi_{1}$} & \colhead{$\theta_{2}$} & \colhead{$\phi_{2}$} & \colhead{$\rperi$}\\
\colhead{} & \colhead{$[\deg]$} & \colhead{$[\deg]$} & \colhead{$[\deg]$} & \colhead{$[\deg]$} & \colhead{[$h^{-1}$ kpc]}}
\startdata
\cutinhead{Disk Orientations}
$b$ & $180$ & $0$ & $0$ & $0$ & $5.0$\\
$c$ & $180$ & $0$ & $180$ & $0$ & $5.0$\\
$d$ & $90$ & $0$ & $0$ & $0$ & $5.0$\\
$e$ & $30$ & $60$ & $-30$ & $45$ & $5.0$\\
$f$ & $60$ & $60$ & $150$ & $0$ & $5.0$\\
$g$ & $150$ & $0$ & $-30$ & $45$ & $5.0$\\
$h$ & $0$ & $0$ & $0$ & $0$ & $5.0$\\
$i$ & $0$ & $0$ & $71$ & $30$ & $5.0$\\
$j$ & $-109$ & $90$ & $71$ & $90$ & $5.0$\\
$k$ & $-109$ & $-30$ & $71$ & $-30$ & $5.0$\\
$l$ & $-109$ & $30$ & $180$ & $0$ & $5.0$\\
$m$ & $0$ & $0$ & $71$ & $90$ & $5.0$\\
$n$ & $-109$ & $-30$ & $71$ & $30$ & $5.0$\\
$o$ & $-109$ & $30$ & $71$ & $-30$ & $5.0$\\
$p$ & $-109$ & $90$ & $180$ & $0$ & $5.0$\\
\cutinhead{Orbital Configurations}
$e1$ & $30$ & $60$ & $-30$ & $45$ & $2.5$\\
$e2$ & $30$ & $60$ & $-30$ & $45$ & $10.0$\\
$e3$ & $30$ & $60$ & $-30$ & $45$ & $15.0$\\
$e4$ & $30$ & $60$ & $-30$ & $45$ & $20.0$\\
$e5$ & $30$ & $60$ & $-30$ & $45$ & $40.0$\\
$e6$ & $30$ & $60$ & $-30$ & $45$ & $30.0$\\
$h1$ & $0$ & $0$ & $0$ & $0$ & $2.5$\\
$h2$ & $0$ & $0$ & $0$ & $0$ & $10.0$\\
$h3$ & $0$ & $0$ & $0$ & $0$ & $15.0$\\
$h4$ & $0$ & $0$ & $0$ & $0$ & $20.0$\\
$h5$ & $0$ & $0$ & $0$ & $0$ & $40.0$\\
$h6$ & $0$ & $0$ & $0$ & $0$ & $30.0$\\
$k1$ & $-109$ & $-30$ & $71$ & $-30$ & $2.5$\\
$k2$ & $-109$ & $-30$ & $71$ & $-30$ & $10.0$\\
$k3$ & $-109$ & $-30$ & $71$ & $-30$ & $15.0$\\
$k4$ & $-109$ & $-30$ & $71$ & $-30$ & $20.0$\\
$k5$ & $-109$ & $-30$ & $71$ & $-30$ & $40.0$\\
$k6$ & $-109$ & $-30$ & $71$ & $-30$ & $30.0$
\enddata
\end{deluxetable}

\subsection{Dissipational Disk Simulations}
\label{subsection:methodology:dissipational}

To gauge the impact of dissipational gas physics on the properties of
merger remnants, we perform a suite of disk galaxy mergers that
include gas cooling, star formation, and supernova feedback.  The
dissipational disk progenitors contain exponential gaseous and stellar
disks and \cite{hernquist1990a} dark matter halos, with their disk
sizes determined by the \cite{mo1998a} formalism as described in \S
\ref{subsection:methodology:dissipationless}.  The vertical structure
of the gaseous disks are determined by an integral constraint from the
surface mass density and the requirement of hydrostatic equilibrium
within the galaxy potential.

The thermal properties of the gas are determined using the multiphase
interstellar medium (ISM) model of \cite{springel2003a}.  Star
formation is prescribed in the manner of \cite{springel2003a},
constrained to approximate the \cite{schmidt1959a} law for disk
galaxies as measured by \cite{kennicutt1998a}, including a density
threshold.  Below this threshold the gas is modeled as a single-phase
medium which is not star-forming.  Dense gas above the threshold is
modeled as a hybrid of cold, dense clouds embedded in a hot, diffuse
medium as envisioned by \cite{mckee1977a}.  The temperature of the hot
phase is set by supernova feedback and the efficiency of cloud
evaporation, and has an energy per unit mass that far exceeds the cold
phase.  Even though most of the gas by mass is cold, the high
temperature of the hot phase more than compensates for its small mass
fraction, acting to pressurize the star-forming gas, and leading to an
effective equation of state $P_{\mathrm{eff}}(\rho)$ that is stiffer
than isothermal \citep[for a numerical fit, see][]{robertson2004a}.  The
multiphase model of \cite{springel2003a} has been generalized by
\cite{springel2005a} to allow for an effective equation of state
parameter $\qEOS$ that linearly interpolates between an isothermal gas
($\qEOS=0$) and the fully-pressurized multiphase ISM model
($\qEOS=1$).  Increasing $\qEOS$ improves the dynamical stability of
the gas and can prevent \cite{toomre1964a} instability even in
gas-rich systems \citep{springel2003a,robertson2004a,springel2005c,robertson2005a}.

For our dissipational models, we re-run the $\Vvir=80-500$ km s$^{-1}$
pure disk merger simulations from \S
\ref{subsection:methodology:dissipationless} with two gas fractions of
$\fgas=0.4,0.8$, each with two equation of state parameters
$\qEOS=0.25,1.0$, at redshifts $z=0$, $2$, $3$, and $6$.  Each
progenitor galaxy has $60,000$ dark matter particles, $40,000$ stellar
disk particles, and $40,000$ gas particles.  The systems are merged on
prograde-prograde parabolic coplanar orbits with $\rperi = 2\rd$.  For
an equation of state parameter $\qEOS=0.25$ we also systematically
vary the gas fraction of $z=0$ progenitors using $\fgas = 0.01$,
$0.025$, $0.05$, $0.1$, $0.2$ and $0.4$.  Our dissipational simulation
category has a total of 164 runs, with the complete list of
simulations provided in Table \ref{table:models}.

\subsection{Full-Model Disk Simulations}
\label{subsection:methodology:full_model}

Our full-model category simulations include the complete physical
model presented in \cite{springel2005a}, accounting for gas cooling,
star formation, supernova feedback, the \cite{springel2003a} ISM
model described in \S \ref{subsection:methodology:dissipational}, and
a prescription for supermassive black hole growth and feedback.  The
supermassive black holes are included as ``sink'' particles, with seed
masses of $10^{5} h^{-1} \Msun$.  The black holes are allowed to grow
according to spherical Bondi-Hoyle-Lyttleton accretion
\citep{hoyle1939a,bondi1944a,bondi1952a}.  The mass accretion rate
$\dot{M}$ is determined from the density and sound speed of the gas
near the black hole. A fraction $\epsilon_{\mathrm{f}} = 0.1$ of the
accretion rate is radiatively released, of which a fraction $\etatherm
= 0.05$ is coupled as thermal feedback into gas within an SPH kernel
smoothing length of the black hole.  The strength of the thermal
coupling is comparable to the thermal feedback coupling of supernova
energy used in cosmological simulations \citep[e.g.][]{abadi2003a},
and reproduces the $\msigma$ relation observed locally
\citep{di_matteo2005a}.  Using the same full-model simulations
presented here, \cite{robertson2005b} have also demonstrated that this
model for black hole accretion and feedback should preserve the
power-law scaling of the $\msigma$ relation between redshifts $z=0-6$.

The full-model simulations augment the pure disk dissipational merger
simulations from \S \ref{subsection:methodology:dissipational} with
supermassive black hole growth as described above.  Each progenitor
galaxy has $40,000$ stellar disk particles and $40,000$ gas particles,
and are merged on prograde-prograde parabolic coplanar orbits with
$\rperi = 2\rd$.  The models are calculated for $\Vvir =80-500$ km
s$^{-1}$ galaxies at $z=0$, $2$, and $3$ and $\Vvir=115-500$ km
s$^{-1}$ galaxies at $z=6$.  Simulations are performed with gas
fractions of $\fgas=0.4, 0.8$ each with equation of state parameters
of $\qEOS=0.25,1.0$. Furthermore, we run 14 variations of the $\Vvir =
160$ km s$^{-1}$ progenitor merger where we change the disk
orientation according to the method of \cite{barnes1992a} to
characterize the effects of disk alignment (see Table
\ref{table:orbits}), and run an additional 18 simulations where for 3
different orientations we vary the pericentric passage distance.  We
also run a set of 5 additional simulations of the $\Vvir = 160$ km
s$^{-1}$ halo where we simulate dark matter halo concentrations of
$\Cvir = 5$, $7$, $9$, $11$, and $13$.  In all, we perform a total of
129 full-model simulations and a complete list of these runs is
provided in Table \ref{table:models}.

\subsection{Spheroid Simulations}
\label{subsection:methodology:spheroids}

To explore the properties of remnants formed by the merging of
spheroidal systems, we also perform a suite of equal mass
spheroid-spheroid mergers.  The spheroid progenitors consist of
stellar spheroids embedded in
dark matter halos, both having 
\cite{hernquist1990a} profiles.  We assume a stellar mass
fraction of $\fstar=0.05$ and the concentrations of the halos are
adjusted to account for the mass-dependence measured in cosmological
simulations (see Section
\ref{subsection:methodology:dissipationless}).  The sizes of the
stellar spheroids are set to follow the \cite{shen2003a} $\re-\Mstar$
relation for massive galaxies as
\begin{equation}
\label{eqn:spheroid_re}
\re = 4.16 \left(\frac{M_{\star}}{10^{11}\Msun}\right)^{0.56} \mathrm{kpc},
\end{equation}
\noindent
\citep[see also][]{boylan-kolchin2005a}.  
We vary the circular velocity of the halos between $\Vvir = 80-500$ km
s$^{-1}$.  The spheroid galaxies merge on either nearly-radial
parabolic orbits with $\rperi = 0.025\Rvir$ or wide elliptical orbits
with ellipticity $\epsilon = 0.5$ and $\rperi = 0.4\Rvir$.  Each
spheroid progenitor has $1,200,000$ dark matter and $80,000$ stellar
particles.  A complete list of the spheroid--spheroid mergers is
provided in Table \ref{table:models}.

To determine the impact subsequent re-merging between ellipticals
formed from disk galaxy mergers might have on the structural
properties of the remnants, we also re-merge disk galaxy remnants from
the $z=0$ full-model simulations.  These remnants are merged on
parabolic orbits with $\rperi = 0.05\Rvir$, where $\Rvir$ is the
virial radius of the original disk progenitor.  We re-merge remnants
with progenitor galaxy circular velocities in the range $\Vvir =
80-226$ km s$^{-1}$.  A complete list of the re-merger simulations is
provided in Table \ref{table:models}.

\subsection{Analysis}
\label{subsection:methodology:analysis}

Each simulation is evolved until the merger is complete and the
remnants are fully relaxed, requiring integrations of typically 2-4
Gyr.  The remnants are then kinematically analyzed by measuring the
half-mass stellar effective radii $\re$, the average one- dimensional
velocity dispersion $\sigmae$ within a circular aperture of radius
$\re$, and the average stellar surface mass density $\Ie$ measured
within $\re$ as $\Ie \equiv \Mstar(r<\re)/ \pi \re^{2}$.  The
quantities $\re$, $\sigmae$, and $\Ie$ are averaged over 100 random
sight lines to the remnant.

Once the FP parameters $\re$, $\sigmae$, and $\Ie$ are determined, we
employ the direct fitting method of \cite{bernardi2003c} to determine
the best-fit FP scalings.  The direct fitting method seeks to minimize
\begin{equation}
\label{eqn:fp_min}
\Delta = \log \re - \alpha \log \sigma - \beta \log \Ie - \delta ,
\end{equation}
\noindent
where $\alpha$ and $\beta$ are the FP scaling indices defined by
Equation (\ref{eqn:fp}).  The minimization of $\Delta$ requires
\begin{equation}
\alpha = \frac{(\sigmaII^2 \sigmaRV^2 - \sigmaIR^2 \sigmaIV^2)}{(\sigmaII^2 \sigmaVV^2 - \sigmaII^4)}
\end{equation}
\begin{equation}
\beta = \frac{(\sigmaVV^2 \sigmaIR^2 - \sigmaRV^2 \sigmaIV^2)}{(\sigmaII^2 \sigmaVV^2 - \sigmaIV^4)}
\end{equation}
\begin{equation}
\delta = \left<\log \re\right> - \alpha \left<\log \sigma\right> - \beta \left<\log \Ie\right> ,
\end{equation}
\noindent
where the average $\left<\log X\right>$ over the $N$ data samples considered is defined as
\begin{equation}
\left<\log X\right> \equiv \sum_i \log X_i / N
\end{equation}
\noindent
and the co-variant dispersion is defined as
\begin{equation}
\sigmaXY^2 = \sum_i \left(\log X_i - \left<\log X\right>\right)\left(\log Y_i - \left<\log Y\right>\right)/N.
\end{equation}
\noindent
The mean-squared scatter about the direct best-fit plane can then be characterized by the quantity
\begin{eqnarray}
\left<\Delta^{2}\right>= \left(\sigmaII^2 \sigmaRR^2 \sigmaVV^2 - \sigmaII^2 \sigmaRV^4 - \sigmaRR^2 \sigmaIV^4\right.\\\nonumber
  \left.- \sigmaVV^2 \sigmaIR^4 + 2 \sigmaIR^2 \sigmaIV^2 \sigmaRV^2\right)\\ \nonumber
	\times\left(\sigmaII^2 \sigmaVV^2 - \sigmaIV^4\right)^{-1}.
\end{eqnarray}
\noindent
When appropriate we compare the quantity $\scatter$ with the 
scatter determined from observational samples.

The remnant properties are also compared to the $\remstar$ relation,
which has been measured observationally in the SDSS \citep{shen2003a}
and may be represented by the power-law form
\begin{equation}
\label{eqn:remstar}
\re \propto \Mstar^{\mu}.
\end{equation}
The $\remstar$ relation allows for a useful comparison of remnant
sizes for simulations with differing angular momenta, progenitor
redshifts, ISM physics, or gas dissipation.  In addition, we find that
the relative location of remnants in the FP can often be related to
the impact of different physical processes on the effective radii.  The
$\remstar$ relation also serves as a useful calibration for our method
to measure the FP properties of the remnants.  As mentioned in \S
\ref{subsection:methodology:spheroids}, a subset of our simulation
suite involves the merging of equilibrium models of spheroids
initialized to satisfy the \cite{shen2003a} relation.  As discussed in
further detail in \S \ref{subsection:results:spheroids}, the analysis
technique used to measure the FP properties of remnants accurately
recovers the \cite{shen2003a} relation when applied to the spheroid
model progenitors and affirms our ability to determine simulated
remnant properties with reasonable fidelity.

\begin{deluxetable}{lccccc}
\tabletypesize{\scriptsize}
\tablecolumns{6} 
\tablewidth{0pt}
\tablecaption{\label{table:scalings}Best-Fit Scalings
}
\tablehead{
\colhead{Models} & \multicolumn{4}{c}{Fundamental Plane} & \colhead{$\remstar$}\\
\colhead{} & \colhead{$\alpha$} & \colhead{$\beta$} & \colhead{$\lambda$} & \colhead{$\left<\Delta^{2}\right>^{1/2}$}& \colhead{$\mu$} }
\startdata
Dissipationless & $2.00$ & $1.01$ & $1.00\pm0.01$ & $0.018$ & $0.45\pm0.03$\\
Dissipational   & $1.58$ & $0.80$ & $0.80\pm0.01$ & $0.065$ & $0.57\pm0.02$\\
Full-Model      & $1.55$ & $0.82$ & $0.79\pm0.01$ & $0.062$ & $0.57\pm0.02$\\
\cutinhead{Gas Fraction $\fgas$ Runs}
$\fgas = 0.01$  & $1.81$ & $0.75$ & $0.97\pm0.01$ & $0.009$ & $0.41\pm0.01$\\
$\fgas = 0.025$ & $2.11$ & $0.74$ & $0.96\pm0.01$ & $0.011$ & $0.41\pm0.01$\\
$\fgas = 0.05$  & $2.07$ & $0.64$ & $0.95\pm0.01$ & $0.011$ & $0.41\pm0.01$\\
$\fgas = 0.1$   & $2.01$ & $0.61$ & $0.92\pm0.01$ & $0.014$ & $0.42\pm0.01$\\
$\fgas = 0.2$   & $1.89$ & $1.20$ & $0.89\pm0.02$ & $0.024$ & $0.44\pm0.01$\\
$\fgas = 0.4$   & $1.64$ & $1.07$ & $0.83\pm0.02$ & $0.033$ & $0.51\pm0.03$\\
\enddata
\end{deluxetable}

Typically the root-mean-squared variation in the velocity 
dispersion and effective radius along different lines-of-sight
to a given remnant are  
$\Delta \sigma \approx 20$ km s$^{-1}$ and 
$\Delta \re \approx 0.5 h^{-1}$ kpc, 
respectively, for remnants formed through mergers of 
$\Vvir = 160$ km s$^{-1}$ progenitors.  The magnitude of
the line-of-sight variation scales with
the velocity dispersion or effective radius, but the 
fractional line-of-sight variation remains similar with 
$\Delta \sigma \approx 0.1\left<\sigma\right>$ and
$\Delta \re \approx 0.1\left<\re\right>$ roughly 
independent of progenitor mass.  The combined quantity
$\sigma^{2}\Ie^{-1}$ has a typical fractional 
line-of-sight variation of 
$\Delta \sigma^{2}\Ie^{-1} \approx 0.3\left<\sigma^{2}\Ie^{-1}\right>$
roughly independent of progenitor mass 
and under this variation galaxies tend to move
along relatively short trajectories in 
the $\re\sigma\Ie$-space that do not deviate far
from the mean FP.  Certain orbits,
such as the head-on orbit considered for spheroidal
mergers in \S \ref{subsection:results:spheroids}, can
induce larger line-of-sight variations but the average
variations reported above are much more frequent in
our remnants.
The impact of these relatively small errors on 
estimates of the power-law correlations of the FP 
and $\remstar$ relations are minor and comparable
to deviations induced by either orbital variations or
the structural properties of progenitors, and
therefore additional
scatter induced by the line-of-sight variations of
individual systems is not considered by
our subsequent analysis.
However, when plotting either the FP or the 
$\remstar$ relations, we provide sample error bars to
show the rough deviations any single galaxy could 
make from their mean positions due to line-of-sight
variations in $\sigma$ or $\re$.

\subsection{Comparison with Observations}
\label{subsection:methodology:comparison}

We adopt the approach of plotting the FP properties of simulated
remnants in the $\re-\sigmae^{2}\Ie^{-1}$ virial plane coordinate
system. The tilt of the FP relative to the virial plane can be
quantified through the power-law relation
\begin{equation}
\label{eqn:tilt}
\re \propto \left(\sigmae^{2}\Ie^{-1}\right)^{\lambda},
\end{equation} 
\noindent
where $\lambda=1$ indicates an alignment of the FP with the virial
plane.  When appropriate, we measure this estimate of the FP tilt
relative to the virial plane by linear least-squares fitting.  Our
choice of coordinate systems is not unique, and alternative
representations of the FP include the $\kappa$-space coordinate system
\citep{bender1992a} or the best-fit FP coordinates determined by
observations in various passbands.  The primary advantage of choosing
the virial plane coordinate system is the easily determined tilt,
which provides a gauge of possible variations in the central 
total-to-stellar mass $\fmtl$ or kinematic nonhomology
of remnants as a function of galaxy mass.

For our purposes, we choose not to use stellar population synthesis to
compare with determinations of the FP in optical passbands.
Observations indicate that trends in the $\mtl$ ratio owing to stellar
population effects (e.g. age or metallicity) as a function of galaxy
mass or luminosity will contribute significantly to the FP tilt,
especially at short-wavelengths 
\citep[for a recent result on this issue, see][]{cappellari2005a}.  
Observations have also determined
that elliptical galaxies are typically old and their stellar
populations redden passively with time
\citep[e.g.][]{bender1996a,van_dokkum1996a}.  These constraints imply 
that to
properly recover the short-wavelength photometric FP, precise
information on the stellar age, metallicity, and formation-redshift
distribution of elliptical galaxies as a function of stellar mass at
$z=0$ must be obtained.  We note that knowing only either the
formation-redshift (e.g. the redshift of the last major gas-rich merger) or
mean stellar age of elliptical galaxies may not be sufficient to
determine the photometric FP of the entire elliptical galaxy
population.  Recent surveys indicate that ellipticals typically
undergo a major dissipationless merger at redshifts $z<1$
\citep{bell2005a,van_dokkum2005a}, and, in principle, such mergers may
induce galaxy mass-dependent tilt from structural effects that are
disjoint from $\mtl$ effects from stellar populations that
characteristically predate those events.  Clearly, applying stellar
population synthesis models to elliptical galaxies produced in
individual galaxy merger simulations to produce simulated photometric FP
scalings without attempting to correct for the
cosmologically-determined properties of the real elliptical population
is likely too naive.  Furthermore, comparing directly stellar-mass FP
scalings determined from simulations with short-wavelength
(e.g. $B$-band or Sloan $g$-band) photometric FP scalings should be
performed with extreme caution as the short-wavelength FP may have
additional sources of tilt not present in the stellar-mass FP.

With these concerns in mind, our stellar-mass FP results will be
compared with the near-infrared (IR) FP determined by
\cite{pahre1998a}.  While, as \cite{pahre1998a} note, tilt owing to
stellar population effects may still be present and is not tightly
constrained, structural or dynamical nonhomologies can contribute
significantly to the FP scaling in the near-IR.  The $K$-band
magnitudes of stellar populations of a given age are less influenced
by metallicity than their optical passband luminosities
\citep[e.g.][]{bruzual2003a}, and typical $K$-band mass-to-light
ratios are within $\approx35\%$ of unity for stellar populations
formed at redshifts $0.75<z<5.0$.  These properties make comparisons
between the stellar-mass and near-IR FP scalings somewhat more
sensible than comparisons with shorter-wavelength photometric FP
scalings, though not ideal.

In principle, a more straight-forward comparison would be to use
high-resolution cosmological simulations of galaxy formation to probe
the fundamental plane with simultaneously accounting for metallicity
and stellar age effects.  Previous attempts to compare results of
cosmological simulations for the scaling laws of elliptical galaxies
with observations have been made \citep[e.g.][]{saiz2004a}, but with
spatial resolution some $\approx20$ times worse than the isolated
merger simulations presented here.  Such spatial resolution is larger
than the effective radii of even moderately-sized
($M_{\mathrm{r}}\geq21)$ early-type galaxies.  However, the
metallicities and ages of the stellar populations of galaxies produced
in cosmological simulations can be estimated throughout their
formation, and the results of stellar population synthesis modeling
would therefore be easy to interpret.

We mention here that semi-analytic techniques could be combined with
the results of high-resolution merger simulations in an attempt to
account for the redshift-dependent formation of the elliptical galaxy
population.  For example, \cite{robertson2005a} used the results of
\cite{hopkins2005a}, who inferred a redshift-dependent black hole mass
function from the quasar luminosity function, to determine the
influence of the redshift-dependent formation times of elliptical
galaxies on the $\msigma$ relation.  Using the results from
\cite{robertson2005a} and \cite{hopkins2005a}, \cite{hopkins2005b}
combined the redshift-dependent properties of galaxy remnants, the
$\msigma$ relation, and the redshift-dependent black hole mass
function to model the evolution of the red-galaxy luminosity function
and color-magnitude relation.  A combination of the results from
\cite{hopkins2005b} and the stellar-mass FP relation presented here
could be used to account for the effects of color-magnitude evolution
on the photometric FP, but would likely involve other assumptions
beyond those employed here and we defer such analysis for future work.

\begin{figure}
\figurenum{1}
\epsscale{1.2}
\plotone{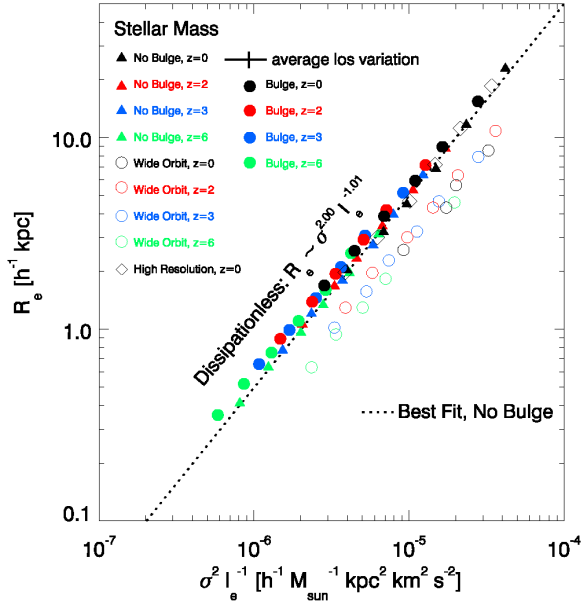}
\caption{\label{fig:fp.dissipationless} 
\small
Fundamental Plane (FP) relation produced by the merging of dissipationless
disk galaxy models appropriate for redshifts $z=0$ (black), $z=2$ (red), 
$z=3$ (blue), and $z=6$ (green) on nearly radial, parabolic orbits.
All models include dark matter halos.
The dissipationless merging of pure disk models (solid triangles) and 
disk models with bulges (solid circles) produce similar FP relations 
nearly parallel to the plane defined by the virial relation.
Increasing the angular momentum of the orbit by lengthening the 
pericentric passage distance of the orbit produces an offset in the FP by 
increasing the 
effective radius of the remnants (open circles), but the systems still 
obtain a
FP scaling similar to the virial plane.
Select higher resolution runs closely follow the FP delineated by their 
lower resolution counter-parts (open diamonds).
For comparison, the best least-squares fit to the FP of pure disk merger 
remnants is plotted (dotted line).
Also shown is the mean deviation induced by line-of-sight variations in
projected quantities for a given remnant (detached error bars).
}
\end{figure}

\begin{figure}
\figurenum{2}
\epsscale{1.2}
\plotone{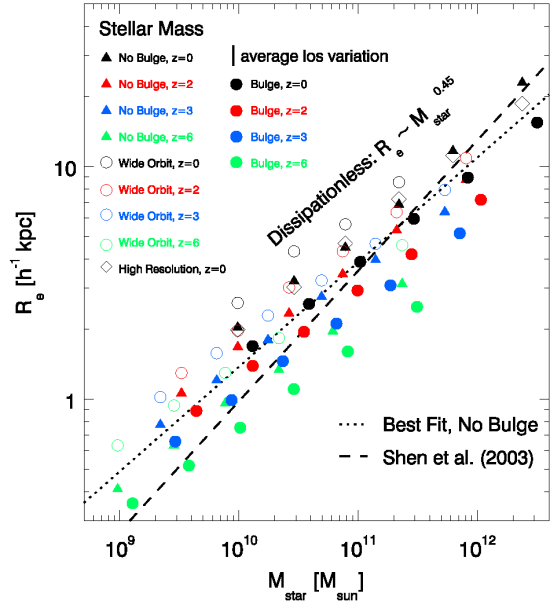}
\caption{\label{fig:re_vs_mstar.dissipationless} 
\small
Effective radius $\re$ -- stellar mass $\Mstar$ relation produced by the 
merging of dissipationless disk galaxy models appropriate for redshifts 
$z=0$ (black), $z=2$ (red), $z=3$ (blue), and $z=6$ (green) on nearly 
radial, parabolic orbits.
All models include dark matter halos.
The dissipationless merging of pure disk models (solid triangles) and 
disk models with bulges (solid circles) produce $\re - \Mstar$ relations 
shallower than that measured for massive galaxies in the Sloan Digital 
Sky Survey \citep{shen2003a}.
For comparison, the best least-squares fit to the $\re - \Mstar$ relation 
of pure disk merger remnants is plotted (dotted line).
Also shown is the mean deviation induced by line-of-sight variations in
projected quantities for a given remnant (detached error bars).
}
\end{figure}
\section{Results}
\label{section:results}

Below, we present the FP and $\remstar$ relations for the
dissipationless, dissipational, full-model, and spheroidal merger
simulations.  For each FP and $\remstar$ relation, we list the best
fit scalings in Table \ref{table:scalings} with 1-$\sigma$
errors on their power-law slopes (i.e. $\lambda$ for the FP 
and $\mu$ for the $\remstar$).

\begin{figure}
\figurenum{3}
\epsscale{1.2}
\plotone{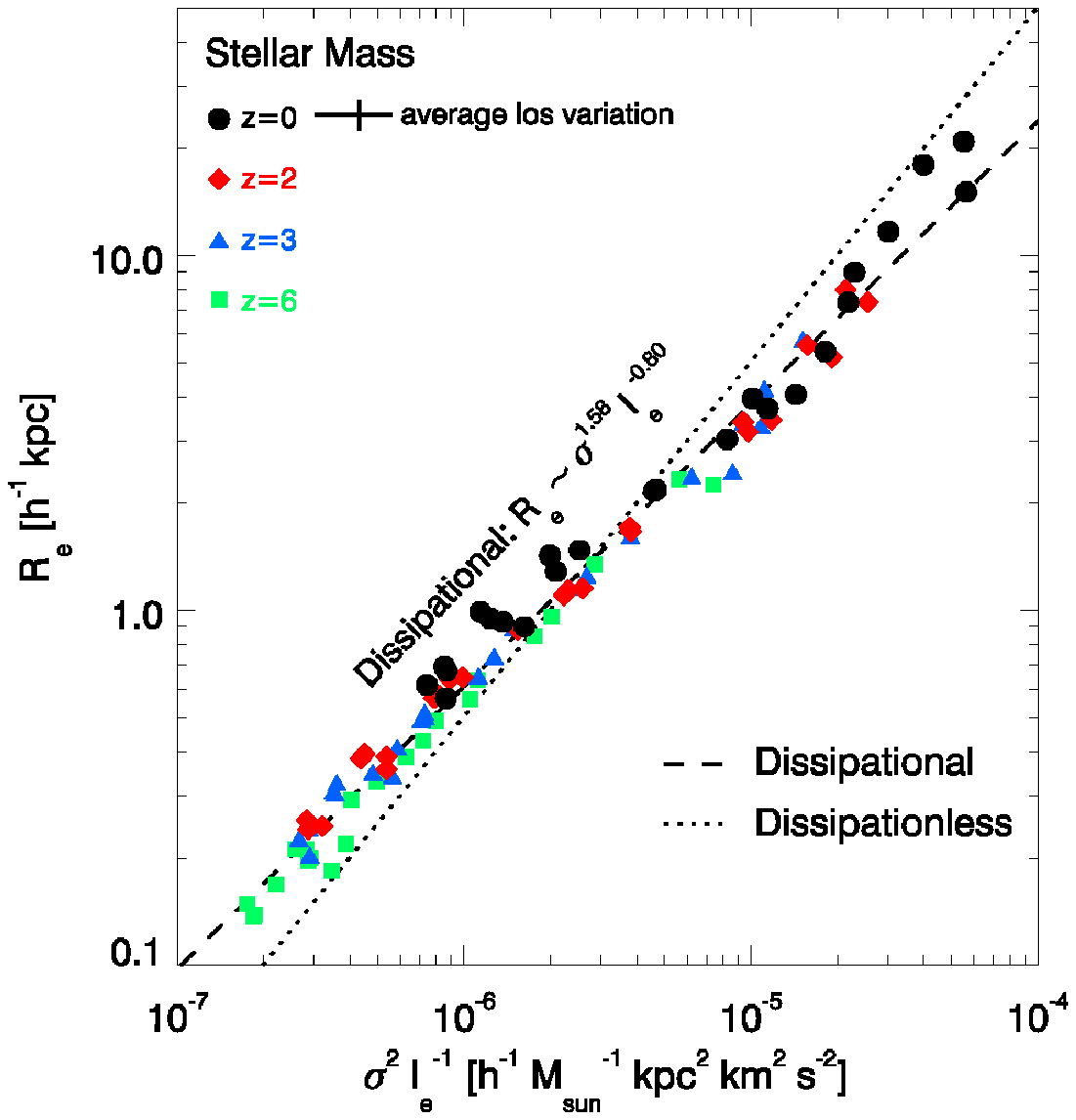}
\caption{\label{fig:fp.dissipational} 
\small
Fundamental Plane (FP) relation produced by the merging of gas-rich disk 
galaxies 
with dark matter halos, star formation and supernova feedback.
Shown are remnants produced by mergers appropriate for redshifts $z=0$ 
(black circles), $z=2$ (red diamonds), $z=3$ (blue triangles), and $z=6$ 
(green squares) with nearly radial, parabolic orbits.
The dissipational merging of pure disk models produces a FP nearly 
parallel to the observed infrared FP \citep{pahre1998a} and is almost 
independent of the redshift scalings of the progenitor systems.
For comparison, the best least-squares fit to the FP delineated by the 
remnants is plotted (solid line).
Also shown is the mean deviation induced by line-of-sight variations in
projected quantities for a given remnant (detached error bars).
}
\end{figure}

\begin{figure}
\figurenum{4}
\epsscale{1.2}
\plotone{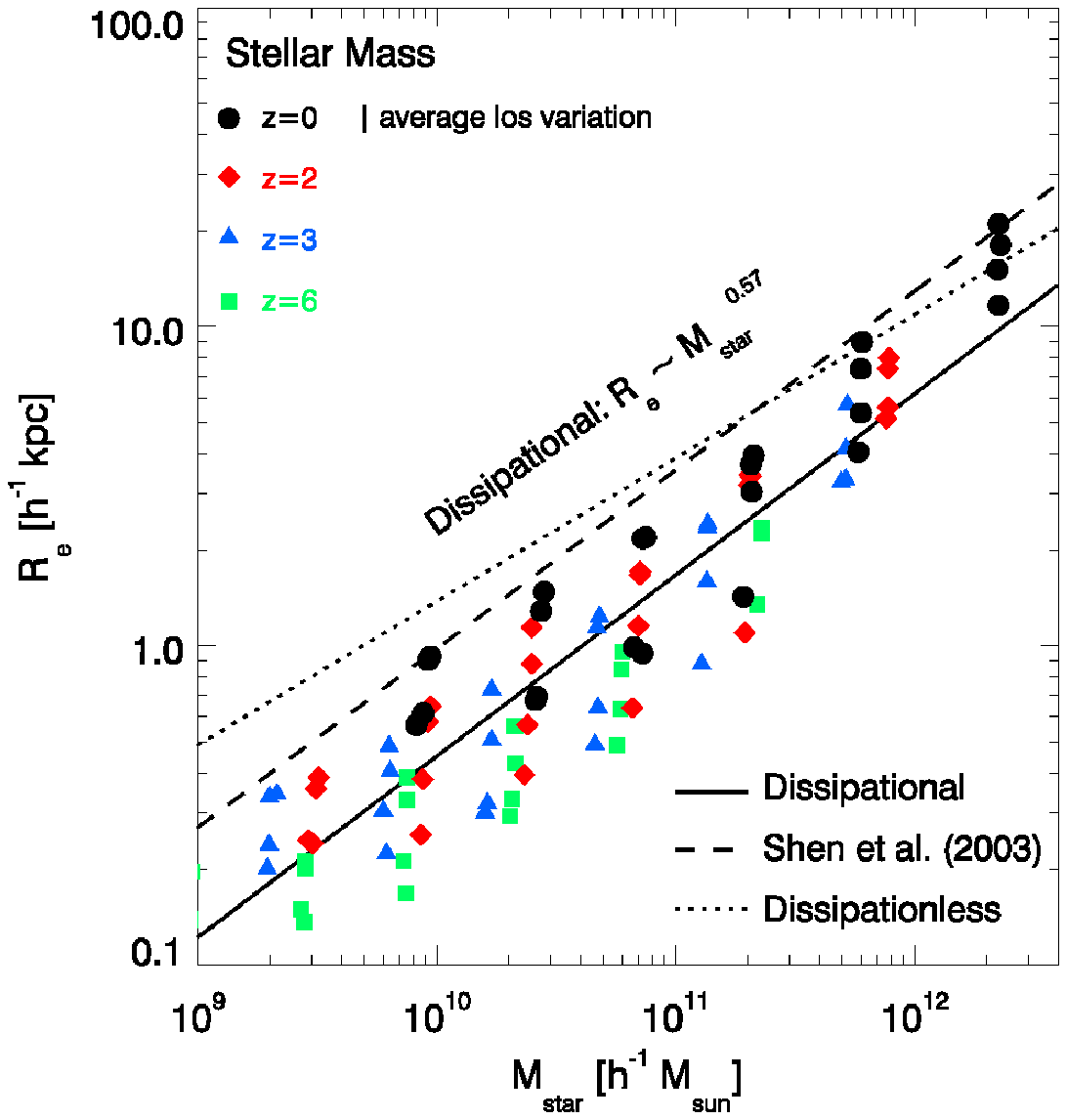}
\caption{\label{fig:re_vs_mstar.dissipational} 
\small
Effective radius $\re$ -- stellar mass $\Mstar$ relation produced by the 
merging of gas-rich disk galaxies with dark matter halos, star formation 
and supernova feedback.
Shown are remnants produced by mergers appropriate for redshifts $z=0$ 
(black circles), $z=2$ (red diamonds), $z=3$ (blue triangles), and $z=6$ 
(green squares) with nearly radial, parabolic orbits.
The dissipational merging of pure disk models produces a $\re - \Mstar$ 
relation roughly parallel to that measured for massive galaxies in the 
Sloan Digital Sky Survey \citep{shen2003a}.
For comparison, the best least-squares fit to the $\re - \Mstar$ 
relations delineated by the dissipational simulations (solid line) and 
dissipationless simulations (dotted line) are plotted.
Also shown is the mean deviation induced by line-of-sight variations in
projected quantities for a given remnant (detached error bars).
}
\end{figure}

\subsection{Dissipationless Disk Simulations}
\label{subsection:results:dissipationless}

The merging of the dissipationless disks described in \S
\ref{subsection:methodology:dissipationless} produces a FP relation
similar to the virial scaling.  Figure \ref{fig:fp.dissipationless}
shows the location of dissipationless merger remnants in the virial
coordinate system, and plots remnants from progenitors appropriate for
various redshifts.  The best-fit FP scalings produced by pure disk
systems (solid triangles) are $\alpha=2.00$, $\beta=1.01$, with almost
no tilt relative to the virial plane ($\lambda=1.00$).  Individually,
progenitors at each of the four simulated redshifts produce merger
remnants also closely aligned with the virial plane
($\lambda=0.97-1.03$).  Including stellar bulges in the progenitors
(solid circles) produces a similar FP scaling ($\alpha=1.95$,
$\beta=0.98$; $\lambda=0.97$).  A more substantial change in the FP is
realized by increasing the pericentric passage distance of the
encounters from $2\rd$ to $0.4\Rvir$ (open circles), which
correspondingly increases the total angular momenta of the merging
systems.  Remnants produced in the dissipationless wide orbit mergers
typically have larger effective radii, inducing an offset of $\Delta
\log \re \sim -0.4$ in the FP.  However, the wide orbit FP scalings
($\alpha=1.97$, $\beta=1.04$; $\lambda=0.97$) are still very similar
to both the nearly radial orbit FP and virial scalings.  Increasing
the number of dark matter particles per halo to $180,000$ in the
progenitors (open diamonds) has very little effect on the resulting
FP, suggesting our results are not strongly influenced by our
numerical resolution.

The $\remstar$ relation produced by merging dissipationless disk
progenitors, shown in Figure \ref{fig:re_vs_mstar.dissipationless},
reflects the trends apparent in the FP those mergers generate and
additional features owing to redshift-dependent progenitor properties.
The merging of pure disk galaxies (solid triangles) appropriate for
various redshifts generates remnants with a shallower mean $\remstar$
relation ($\mu\approx0.45$, solid line) than that measured in
late-type galaxies in the Sloan Digital Sky Survey
\citep[][$\mu\approx0.56$]{shen2003a}.  The remnants also
systematically decrease in effective radius with redshift, reflecting
the decrease in progenitor disk scalelength and pericentric passage
distance.  Wider orbit mergers (open circles) with larger angular
momenta produce remnants with larger effective radii while galaxies
containing bulges (solid circles), and therefore less specific angular
momenta, produce remnants with smaller effective radii.  These results
are consistent with expectations for dissipationless systems
where the energy and angular momenta of the model system are
manifestly conserved.  As a final note for the dissipationless runs,
increasing the numerical resolution of the dark matter by $3\times$
(open diamonds) gives consistent results, suggesting that artificial
heating of the stellar component by discreteness effects in the dark
matter halo is not strongly influencing the structure of the
remnants.

\begin{figure}
\figurenum{5}
\epsscale{1.2}
\plotone{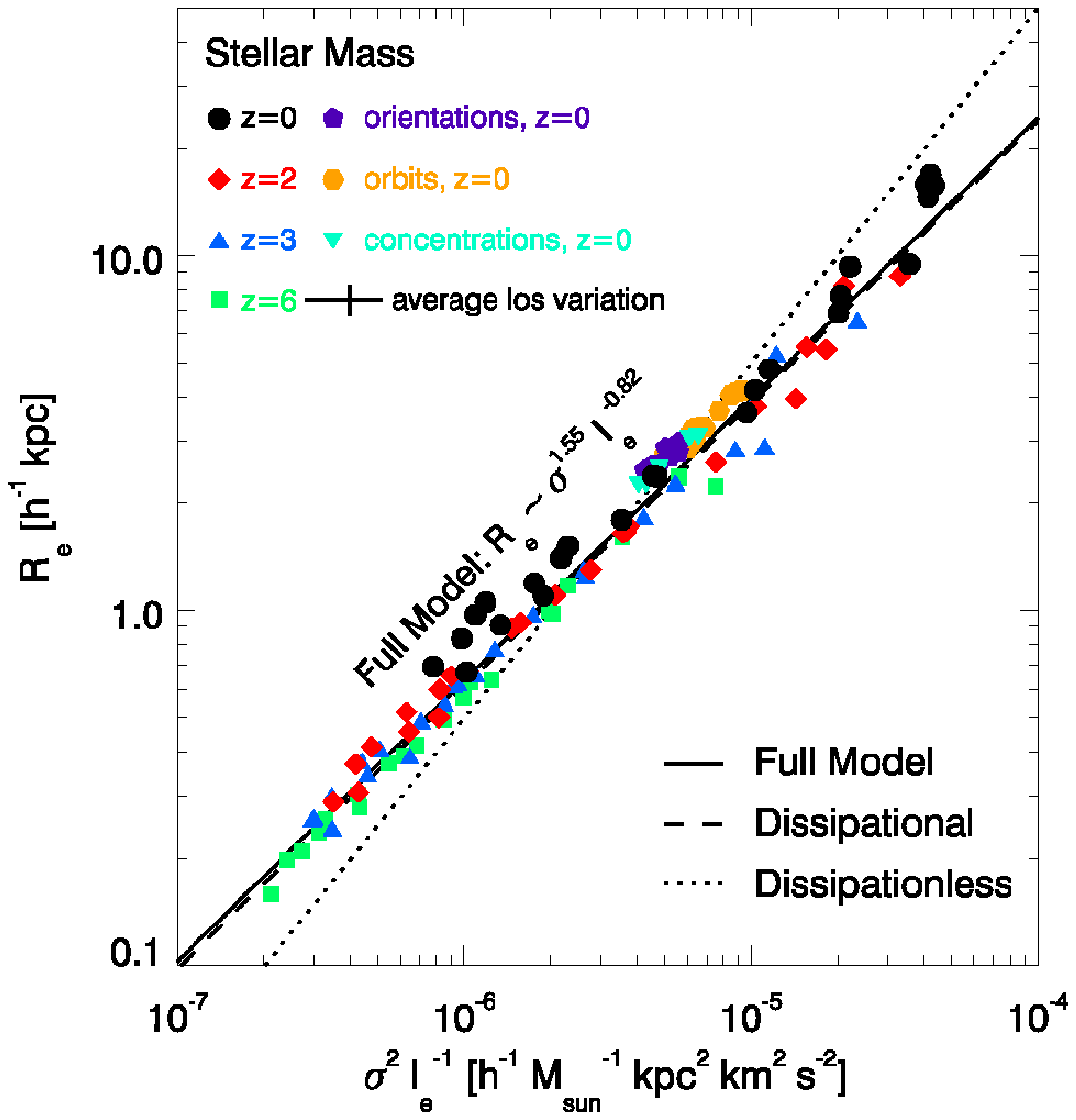}
\caption{\label{fig:fp} 
\small
Fundamental Plane (FP) relation produced by the merging of gas-rich disk 
galaxies with dark matter halos, star formation, supernova feedback, and 
a prescription for feedback from accreting supermassive black holes.
Shown are remnants produced by mergers appropriate for redshifts $z=0$ 
(black circles), $z=2$ (red diamonds), $z=3$ (blue triangles), and $z=6$ 
(green squares) with nearly radial, parabolic orbits.
The merging of pure disk galaxies using our full physical model produces 
a FP nearly parallel to the observed infrared FP \citep{pahre1998a} and 
nearly coincident with the FP produced by similar simulations without 
black holes.
The FP relation is roughly independent of the redshift scalings of the 
progenitor systems and the location of remnants within FP is fairly 
insensitive to a large variety of disk orientations (purple pentagons)
and orbital configurations (orange hexagons), as changes in the effective 
radius are compensated by changes in the velocity dispersion and surface 
mass density.
For comparison, the best least-squares fit to the FP delineated by the 
remnants is plotted (solid line).
Also shown is the mean deviation induced by line-of-sight variations in
projected quantities for a given remnant (detached error bars).
}
\end{figure}

\begin{figure}
\figurenum{6}
\epsscale{1.2}
\plotone{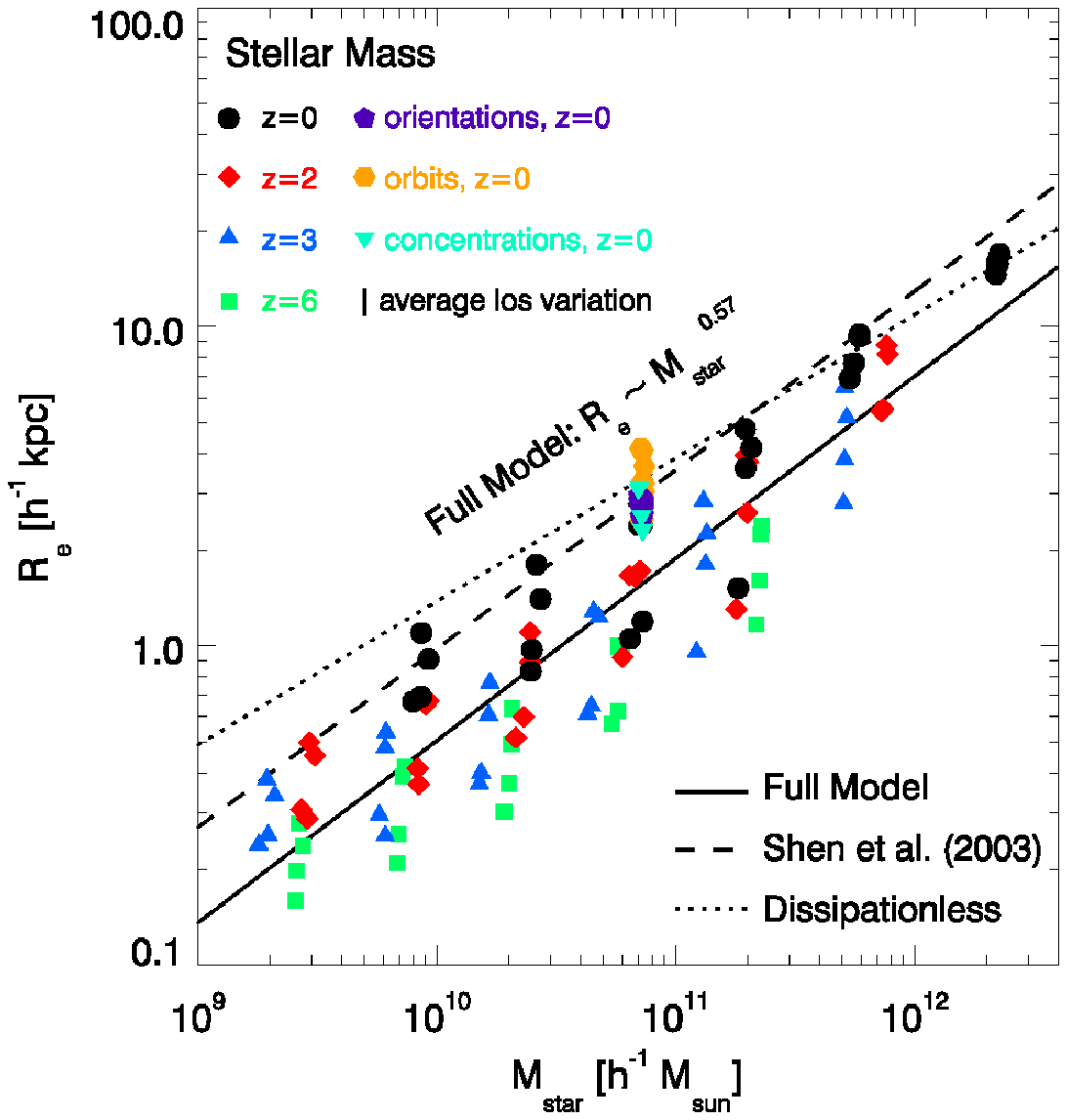}
\caption{\label{fig:re_vs_mstar}
\small
Effective radius $\re$ -- stellar mass $\Mstar$ relation produced by the 
merging of gas-rich disk galaxies with dark matter halos, star formation, 
supernova feedback, and a prescription for feedback from accreting 
supermassive black holes.
Shown are remnants produced by mergers appropriate for redshifts $z=0$ 
(black circles), $z=2$ (red diamonds), $z=3$ (blue triangles), and $z=6$ 
(green squares) with nearly radial, parabolic orbits.
Also plotted are remnants produced by varying the system angular momentum 
through the initial disk orientation
(purple pentagons) and pericentric passage distances (orange hexagons) 
for a single pair of progenitor models, which produces a spread in the 
remnant effective radius.
The merging of pure disk galaxies using our full physical model produces a
$\re - \Mstar$ relation roughly parallel to that measured for massive 
galaxies in the Sloan Digital Sky Survey \citep{shen2003a}, though with 
an offset.
The resultant $\re - \Mstar$ relation shifts downward with the redshift 
of the progenitor systems as the remnants decrease with size.
For comparison, the best least-squares fit to the $\re - \Mstar$ relation 
delineated by these remnants (solid line), as well as the dissipationless 
disk model $\re - \Mstar$ relation (dotted line) and the \cite{shen2003a} 
relation (dashed line) are plotted.
Also shown is the mean deviation induced by line-of-sight variations in
projected quantities for a given remnant (detached error bars).
}
\end{figure}

\subsection{Dissipational Disk Simulations}
\label{subsection:results:dissipational}

The merging of gas-rich disk galaxy progenitors including dissipation
results in FP and $\remstar$
relations that differ substantially from the analogous relations
produced by the merging of dissipationless disks.  Figure
\ref{fig:fp.dissipational} shows the FP relation generated by the
dissipational model disk galaxies appropriately scaled for various
redshifts.  The dissipational model FP displays a tilt relative to the
virial plane ($\lambda=0.8$), with a scaling ($\alpha=1.58$,
$\beta=0.80$) similar to the near-IR photometric FP
\citep[][$\alpha=1.53$, $\beta=0.79$]{pahre1998a}.  The dissipational
model FP includes remnants produced from progenitors that vary by a
factor $2\times$ in gas fraction ($\fgas=0.4,0.8$), including either
strongly ($\qEOS=1.0$) or weakly ($\qEOS=0.25$) pressurized ISM
equations-of-state.  For gas-rich systems ($\fgas\geq0.4$), the gas
fraction and ISM physics of the progenitors have only a slight effect
on the FP.  The redshift-dependent properties of progenitor systems
have little effect on the simulated FP plane, even as the structural
properties of the disk galaxies vary substantially with redshift.
Combined, varying the gas fraction above $\fgas>0.4$, changing the
ISM pressurization dramatically, and scaling the progenitor systems
for redshifts $z=0-6$ produce only a small amount of scatter in the
dissipational model FP ($\scatter=0.065$).

The $\remstar$ relation of the remnants, shown in Figure
\ref{fig:re_vs_mstar.dissipational}, is more strongly influenced by
the range of progenitor properties.  The smaller progenitor galaxies
appropriate for higher redshifts produce smaller remnants, while the
less-pressurized ISM models also decrease the effective radii of the
remnants.  The dissipational model $\remstar$ relation is
significantly steeper (solid line, $\mu=0.57$) than the relation
produced by the dissipationless merging of disk galaxies (dotted line,
$\mu=0.45$), and compares well with the relation measured for
late-type galaxies in the SDSS \citep{shen2003a}.  The dissipational
simulations produce an $\remstar$ relation that has a lower $\re$
normalization than the \cite{shen2003a} relation, but subsequent
re-merging and a cosmologically-representative distribution of orbits
will likely decrease the discrepancy by increasing $\re$ at a given
stellar mass.  Since these the dissipationless remnants lie above the
\cite{shen2003a} relation, re-merging and higher angular momentum
orbits will only further increase this discrepancy with real
elliptical galaxies.  While a proper accounting for the distribution
of formation redshifts for elliptical galaxies would alter the
$\remstar$ relation by weighting the remnants unequally, the mergers
of disk galaxies where dissipation is important will nevertheless
produce a $\remstar$ relation that is steeper than that produced by
dissipationless merging.

\subsection{Full-Model Disk Simulations}
\label{subsection:results:full_model}

Introducing the effects of black hole feedback through the full-model
simulations produces a set of fundamental scaling relations similar to
those in dissipational mergers without black holes.  The fundamental
plane produced by the full-model simulations yields nearly the same FP
scalings ($\alpha=1.55$, $\beta=0.82$)
and tilt ($\lambda=0.79$) as the dissipational simulations (see \S
\ref{subsection:results:dissipational}),
and is similar to the observed near-IR scalings \citep{pahre1998a}.
These simulated remnants exhibit a scatter about their mean FP of 
$\scatter = 0.062$, comparable to both the observed scatter in the FP at
optical wavelengths \citep[e.g.][]{bernardi2003c} and that produced
by the dissipational simulations.
Black hole feedback causes the full-model remnants to be slightly
larger than the dissipational model remnants as feedback-driven winds
remove gaseous material from the innermost regions of the remnants
that would otherwise contribute to the central stellar content.
Figure \ref{fig:fp} shows the full-model remnants produced by
mergers appropriate for various redshifts on nearly radial, parabolic
orbits.  The resultant FP relation is roughly independent of the
redshift scalings of the progenitor systems and the location of
remnants within the FP is fairly insensitive to the orientations of
the disks (purple pentagons, see Table \ref{table:orbits}) and orbital
configuration (orange hexagons, see Table \ref{table:orbits}).  As
these simulations indicate, varying the total angular momentum
by changing either the orientation of the disks (spin) or the
pericentric passage distance (orbital) does not strongly influence the
FP of the remnants.  Typically the angular momentum of the orbit only
influences the effective radii of the remnants, with the velocity
dispersion and surface mass density changing in a compensating fashion
to maintain the FP scalings.  In addition, varying the concentration
of the dark matter halo by a factor of $\sim 2.5$ does little to
change the FP scalings (cyan triangles).

The influence of angular momentum on the remnant properties is plainly
visible in the $\remstar$ relation for the full-model simulations
(Figure \ref{fig:re_vs_mstar}).  As is evident in the FP scalings,
simulated mergers appropriate for redshifts $z=0-6$ produce results
very similar to the dissipational runs with a comparable $\remstar$
scaling ($\mu=0.57$) and slightly larger remnants.  However, varying
the angular momentum of the merging system through changing the
progenitor disk orientations (purple pentagons) and pericentric
passage distances for several orientations (orange hexagons) can as
much as double the average effective radius of the remnant produced by
the same progenitors.  The dark matter halo concentration also has a
noticeable effect on the galaxy size (cyan triangles).  We reiterate
that these changes in the effective radius only induce a spread
parallel to the FP (see above).  Moreover, as is the case for the
dissipational simulations without black hole growth, the $\remstar$
relation for the full model remnants would only benefit from
subsequent re-merging or more cosmologically representative orbits by
increasing $\re$.

\begin{figure}
\figurenum{7}
\epsscale{1.2}
\plotone{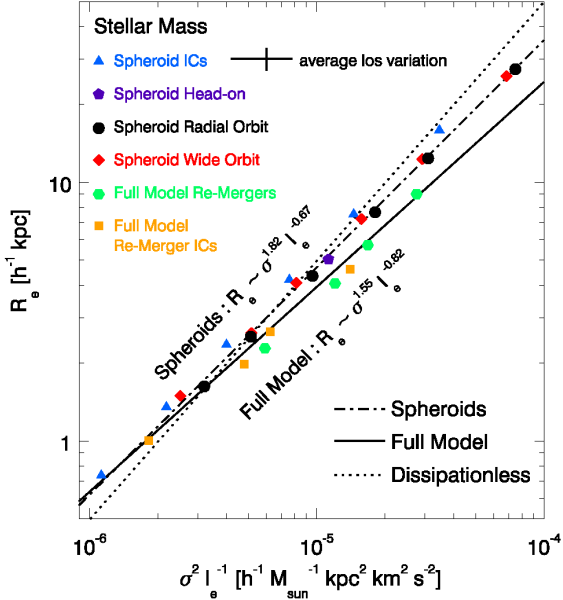}
\caption{\label{fig:fp.spheroids} 
\small
Fundamental Plane (FP) relation produced by the merging of spheroidal 
galaxy models.
Using \cite{hernquist1990a} stellar spheroids models with dark matter 
halos as initial conditions (blue triangles), 
the remnants produced by dissipationless spheroid mergers appropriate for 
redshift $z=0$, with nearly radial, parabolic orbits (black circles) and 
wide, elliptical orbits with circularity $\epsilon = 0.5$ (red diamonds) 
are calculated.
The remnants of spheroidal mergers produce a FP relation similar to their 
progenitor systems, roughly independent of the orbital energy or angular 
momentum.
Also shown are select spheroidal remnants from gas-rich disk galaxy merger
simulations that include star formation, supernova feedback, and a 
prescription for feedback from accreting supermassive black holes 
(orange squares).
The disk galaxy merger remnants occupy a FP relation similar to that 
observed in infrared observations (solid line).
The re-merging of these disk galaxy remnants on nearly radial, parabolic 
orbits (green circles), further demonstrates that the merging of 
spheroidal remnants produces a FP similar to that occupied by the 
progenitor systems
\citep[e.g.][]{capelato1995a,dantas2003a,nipoti2003a,boylan-kolchin2005a}.
For comparison, the best least-squares fit to the FP delineated by the 
dissipationless disk merger remnants (dotted line) and spheroidal merger 
remnants (dash-dotted line) is plotted.
Also shown is the mean deviation induced by line-of-sight variations in
projected quantities for a given remnant (detached error bars).
}
\end{figure}

\begin{figure}
\figurenum{8}
\epsscale{1.2}
\plotone{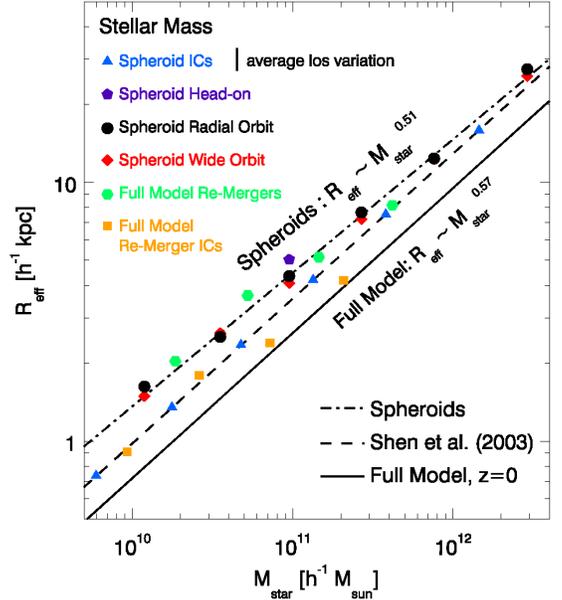}
\caption{\label{fig:re_vs_mstar.spheroids} 
\small
Effective radius $\re$ -- stellar mass $\Mstar$ relations produced by the 
merging of spheroidal galaxy models.
The mergers of \cite{hernquist1990a} stellar spheroids initial conditions 
with dark matter halos (blue triangles), initialized to the 
\cite{shen2003a} relation (dashed line), with nearly radial, parabolic 
orbits (black circles) and wide, elliptical orbits with circularity 
$\epsilon = 0.5$ (red diamonds) are simulated.
The remnants of spheroidal mergers produce a $\remstar$ relation slightly 
shallower than the observed relation, roughly independent of the orbital 
energy or angular momentum.
Also shown are spheroidal remnants from gas-rich disk galaxy merger 
simulations that include star formation, supernova feedback, and a 
prescription for feedback from accreting supermassive black holes 
(orange squares).
The disk galaxy merger remnants occupy a $\remstar$ relation with the 
same 
scaling as observed in SDSS (dashed line).
The re-merging of these disk galaxy merger remnants on nearly radial, 
parabolic orbits (green circles) produces a similar $\re - \Mstar$ 
relation as the dissipationless merging of spheroidal galaxies.
For comparison, the best least-squares fit to the $\re - \Mstar$ relation 
delineated by the dissipationless spheroidal merger remnants 
(dashed-dotted line),
the \cite{shen2003a} relation, and the relation produced by simulations 
using the full physical model (solid line) are plotted.
Also shown is the mean deviation induced by line-of-sight variations in
projected quantities for a given remnant (detached error bars).
}
\end{figure}

\subsection{Spheroid Simulations}
\label{subsection:results:spheroids}

As demonstrated in \S \ref{subsection:results:dissipationless}--
\ref{subsection:results:full_model}, the merging of dissipationless
disk galaxies produces a FP with almost no tilt relative to the
virial plane and a $\remstar$ relation that is shallower than observed,
while gas-rich disk galaxies yield FP relations with
substantial tilt and reasonable $\remstar$ scalings.  A natural
extension of these calculations is to test whether the subsequent
merging of the remnants or model galaxies similar in structure to the
remnants preserves the FP and $\remstar$ relations produced by the
first generation of mergers.  The spheroidal galaxy model progenitors
described in \S \ref{subsection:methodology:spheroids} that are
initialized to obey the observed \cite{shen2003a} $\remstar$ relation
delineate a tight FP relation (blue triangles, Figure
\ref{fig:fp.spheroids}).  Merging these spheroidal galaxy models on
either nearly-radial parabolic orbits (red diamonds), wide
slightly-bound orbits (black circles), or in an exactly head-on encounter
(purple pentagon) produces a similar tight FP relation, only slightly
tilted relative to the progenitor FP scalings.  As described
previously, altering the angular momentum of the orbits has the effect
of increasing the effective radii of the remnants but does not
strongly affect the FP relation as the velocity dispersion and surface
mass density compensate to roughly maintain the progenitor FP
scalings.

A similar result is obtained by re-merging remnants from the
full-model simulations.  The FP delineated by the full-model remnants
(Figure \ref{fig:fp.spheroids}, solid line) is roughly recovered by
re-merging the disk galaxy remnants, which demonstrates that the
approximate preservation of the FP relation holds for relations
instilled by hand (as for spheroidal model galaxies) and those
generated self-consistently through the merging process (as for the
full-model merger remnants).  The re-merger remnants are consistently
larger in effective radii when compared with their progenitors, also
in agreement with the spheroidal galaxy model mergers and previous
work \citep[e.g.][]{boylan-kolchin2005a}.

The behavior of the $\remstar$ relation under merging demonstrates
the increase of $\re$ of spheroidal and re-merger remnants
relative to their progenitor properties.  As noted previously
\citep{barnes1992a,hernquist1993b,cole2000a,nipoti2003a,shen2003a,boylan-kolchin2005a},
the effective radius of the spheroidal merger remnants approximately
doubles in an equal-mass merger and will introduce scatter parallel
to the $\re$-axis.  However, as is apparent from Figure
\ref{fig:re_vs_mstar.spheroids}, the $\remstar$ relation only
moderately changes as a mass-sequence (e.g. $\mu=0.56$ for spheroidal
progenitors, $\mu=0.52$ for spheroidal merger remnants).  
Re-mergers of disk galaxy remnants produce a similar change in the
$\remstar$ relation.  A dissipationless merger between spheroids then
is expected to introduce scatter into the $\remstar$ relation but does
not completely destroy the $\remstar$ correlation of the
progenitors.  The presence of a well-defined $\remstar$ correlation at
$z=0$ is therefore not a strong constraint on the dissipationless
merging of spheroids \citep[c.f.][]{shen2003a}, although substantial
additional merging beyond the recently inferred $0.5-1$
dissipationless mergers at redshifts $z<1$ typical for spheroidal
galaxies \citep{bell2005a,van_dokkum2005a} could possibly induce
noticeable scatter into the $\remstar$ relation beyond that reported
by \cite{shen2003a}.

\begin{figure}
\figurenum{9}
\epsscale{1.2}
\plotone{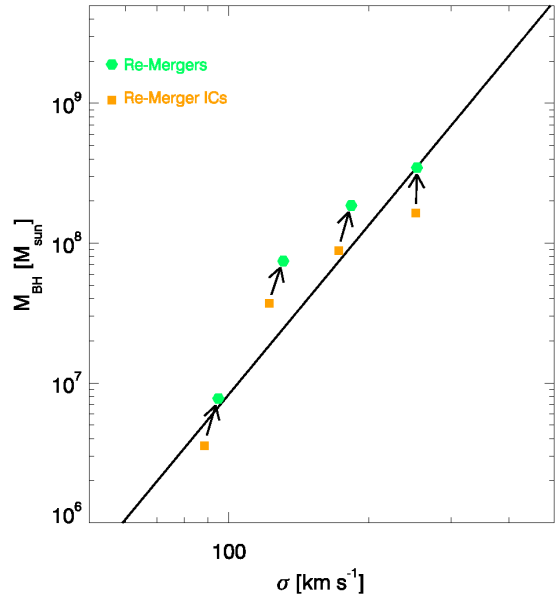}
\caption{\label{fig:remergers.msigma}
\small
Black hole mass $\MBH$ -- stellar velocity dispersion relation $\sigma$ 
relation measurements for galaxies produced by the re-merging of disk 
galaxy merger remnants that obey the local $\msigma$ relation 
\citep[e.g.,][solid line]{tremaine2002a}.  While the number of re-mergers
examined is limited, a single generation of dissipationless mergers after
the initial formative gas-rich mergers that generate the $\msigma$ 
relation
is not expected to strongly alter the observed $\msigma$ relation but may
be a source of scatter.
}
\end{figure}

We note briefly here that the maintenance of a $\remstar$ relation
with $\mu\approx0.5$ also will keep the $\msigma$ relation scaling
of $\MBH \propto \sigma^{4}$, while possibly inducing some additional
scatter owing to a shift in the normalization toward lower velocity
dispersions at a given black hole mass.  The velocity dispersion of a
spheroidal galaxy is roughly
\begin{equation}
\label{eqn:msigma.argument.1}
\sigma^{2} \propto \frac{\Mstar}{\re}.
\end{equation}
\noindent
The maintenance of the $\remstar$ relation with $\mu\approx0.52$ after
re-merging and the assumption that the stellar mass $\Mstar$ and black
hole mass $\MBH$ both double then implies
\begin{equation}
\label{eqn:msigma.argument.2}
\MBH \propto \sigma^{4.2},
\end{equation}
\noindent
which is approximately the observed $\msigma$ scaling at $z=0$.
Correspondingly, the $\msigma$ relation of the re-merger
remnants that include black holes (see Figure
\ref{fig:remergers.msigma}) does not differ strongly from the
$\msigma$ relation delineated by their progenitor systems.  While
$\sigma$ does not increase dramatically enough to perfectly preserve
the observed relation \citep{nipoti2003a}, the measured $\remstar$
relation of the re-merger remnants provides a preliminary explanation
for the rough maintenance of the $\msigma$ relation through infrequent
dissipationless mergers.  We also note that if dissipationless major
merging happens more frequently for more massive systems, then the 
$\msigma$ relation may also steepen relative to the
\cite{tremaine2002a} best-fit relation at higher masses.  
A signature of such dynamical reprocessing may appear as a mass-dependent
scatter in the local $\msigma$ relation \citep{robertson2005b}.

\section{The Impact of Progenitor Gas Fraction on Merger Remnant Properties}
\label{section:gas_fraction}

The results presented in \S \ref{section:results} demonstrate that the
merger of gas-rich systems produces a substantially different
fundamental plane scaling than the virial scalings obtained by the
merging of dissipationless disk galaxies.  However, the origin of the
FP tilt measured for the simulations that include the effects of
dissipational physics has not yet been illuminated.  To better
demonstrate that dissipational physics does indeed produce a FP tilt
for sufficiently gas-rich systems and to demonstrate the manner by
which gas dissipation gives rise to FP tilt, we now present the
results of a sequence of dissipational model mergers in which the gas
fraction $\fgas$ of the progenitor disks has been varied
systematically from $\fgas=0.01$ to $\fgas=0.4$.  The details of these
simulations are described in \S
\ref{subsection:methodology:dissipational} and listed in Table
\ref{table:models}.

\subsection{Fundamental Scaling Relations}
\label{subsection:gas_fraction:scalings}

\begin{figure}
\figurenum{10}
\epsscale{1.2}
\plotone{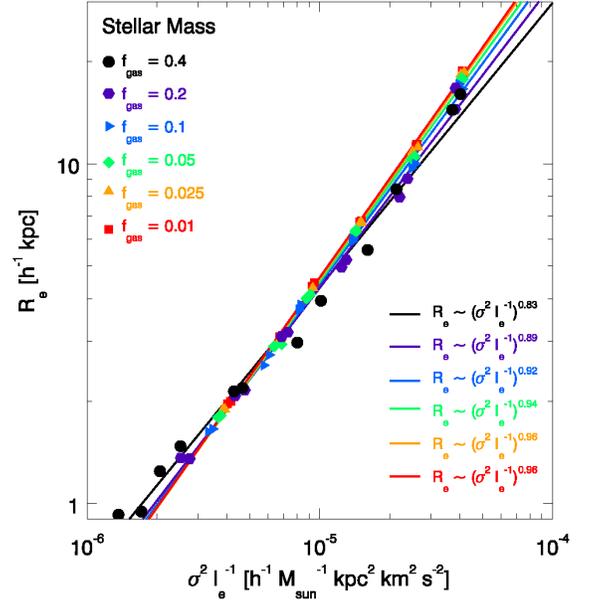}
\caption{\label{fig:fp.gf}
\small
Fundamental Plane (FP) relation produced by the merging of disk galaxies
with dark matter halos, star formation and supernova feedback as a 
function of progenitor gas content.
The galaxy models are appropriate for $z=0$ and are merged on nearly 
radial, parabolic orbits.
Shown are remnants produced by progenitor systems with gas fraction 
$\fgas = 0.01$ (red squares and line), $\fgas = 0.025$ (yellow triangles 
and line), $\fgas = 0.04$ (green diamonds and line), $\fgas = 0.1$ (blue 
triangles and line), $\fgas = 0.2$ (purple hexagons and line), and 
$\fgas = 0.4$ (black circles and line).
As the gas fraction of the progenitor systems increases past 
$\fgas = 20\%$, the dissipational effects of gas cooling and star 
formation induce a substantial tilt in the resulting FP relation.
Nearly dissipationless progenitor ($\fgas < 20\%$) produce FP relations 
roughly parallel to the plane delineated by the virial relation, while 
for systems where gas dissipation is important the resultant FP is 
similar to the observed infrared FP \citep{pahre1998a}.
The increasing tilt is generated through a decrease in the remnant 
effective radius with increasing gas content and the corresponding change 
in the central stellar mass fraction (see also Figures 
\ref{fig:re_vs_mstar.gf} -- \ref{fig:mtl.vc6}).
}
\end{figure}

The Fundamental Plane relation produced by the merging of disk
galaxies models with dissipation as a function of progenitor gas
content is shown in Figure \ref{fig:fp.gf}.  The remnants produced
by progenitor systems with various gas fractions, $\fgas$ are plotted.
The impact of dissipation on the FP tilt of remnants exhibits a
behavior where for remnants of progenitors with $\fgas<0.2$
little FP tilt is induced by gas dissipation while for remnants of
progenitors with $\fgas>0.2$ substantial FP tilt is induced (see Table
\ref{table:scalings} for the best-fit scalings).  The
nearly-dissipationless simulations ($\fgas<0.1$) produce FP scalings
similar to the virial scalings ($\lambda=0.92-0.97$), while the
gas-rich mergers ($\fgas=0.4$) produce a substantial tilt
($\lambda=0.83$).

\begin{figure}
\figurenum{11}
\epsscale{1.2}
\plotone{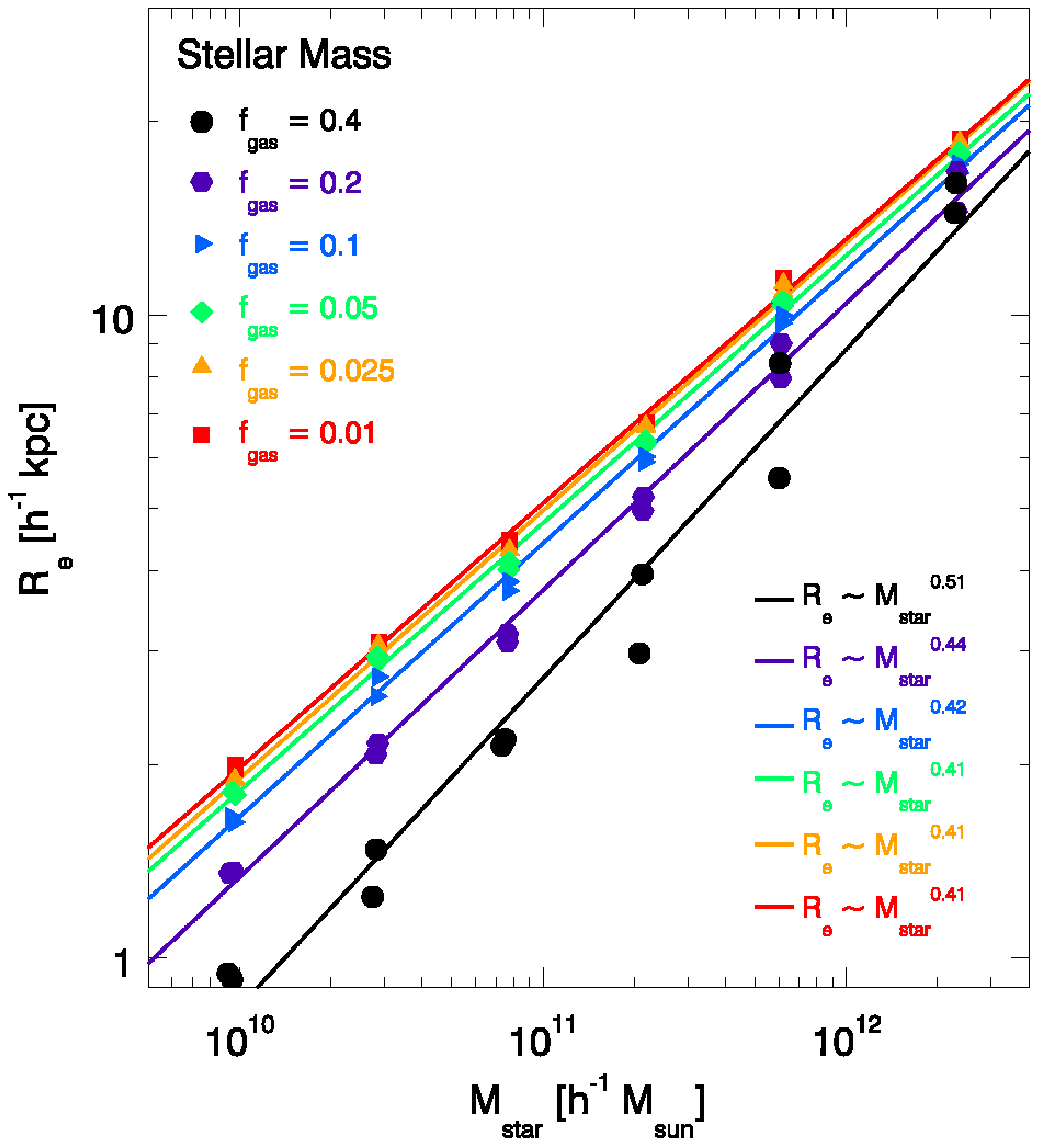}
\caption{\label{fig:re_vs_mstar.gf}
\small
Effective radius $\re$ -- stellar mass $\Mstar$ relation produced by the 
merging of gas-rich disk galaxies with dark matter halos, star formation 
and supernova feedback as a function of progenitor gas content.
The galaxy models are appropriate for $z=0$ and are merged on nearly 
radial, parabolic orbits.
Shown are remnants produced by progenitor systems with gas fraction 
$\fgas = 0.01$ (red squares and line), $\fgas = 0.025$ (yellow triangles 
and line), $\fgas = 0.04$ (green diamonds and line), $\fgas = 0.1$ (blue 
triangles and line), $\fgas = 0.2$ (purple hexagons and line), and 
$\fgas = 0.4$ (black circles and line).
As the gas fraction of the progenitor systems increases, the dissipational
effects of gas cooling and star formation produce smaller remnant 
effective radii and a corresponding change in the central stellar mass 
fraction (see also Figures \ref{fig:fp.gf}, \ref{fig:mtl.vc2}, and 
\ref{fig:mtl.vc6}).
These dissipational effects generate a substantial tilt in the 
Fundamental Plane above a gas fraction of roughly $\fgas > 30\%$.
}
\end{figure}

The $\remstar$ relation of the remnants also displays a 
behavior where above a critical gas fraction $\fgas\approx0.3$ 
dissipation is important for determining the remnant properties.
Below $\fgas\approx0.2$ the $\remstar$ relation is shallower than the
observed \cite{shen2003a} relation and is similar to the $\remstar$
relation produced by dissipationless merging of purely stellar disks.
As the gas fraction of the progenitor systems increases beyond
$\fgas=0.2$, the dissipational effects of gas cooling and star
formation produce smaller remnant effective radii and a steeper
$\remstar$ relation (see Figure \ref{fig:re_vs_mstar.gf}).  The
decrease in effective radius is most dramatic in lower-mass systems,
where $\re$ can decrease by as much as $50\%$.  The tilt induced in
the FP relation with progenitor gas fraction by the increasing
importance of gas dissipation appears closely related to this
mass-dependent decrease in effective radius, as the velocity
dispersion and surface mass density adjust with the change in
effective radius to produce FP tilt as the dissipative effects
increase.  We note here that the equation of state of the ISM also
influences the effective radii, with the less-pressurized ISM model
leading to smaller remnants as dissipation becomes more efficient with
decreasing ISM pressurization.

\subsection{Phase-Space Density}
\label{subsection:gas_fraction:psd}

While the behavior of the fundamental scaling relations of the merger
remnants as a function of progenitor gas fractions demonstrates that
dissipation can induce FP tilt by steepening the $\remstar$
relation, these structural parameter correlations reflect the
composition of more fundamental properties of the remnants.
One such property is the fine-grained
phase-space density $f(x,v)$ at the spatial coordinate $x$ and
velocity coordinate $v$. The quantity $f(x,v)\mathrm{d}x\mathrm{d}v$
describes the probability of a stellar mass element $\mathrm{d}\Mstar$
being located within the differential phase-space volume
$\mathrm{d}x\mathrm{d}v$ centered at coordinates $(x,v)$.  For a
purely collisionless system, $f(x,v)$ satisfies the Boltzmann equation
\begin{equation}
\label{eqn:cbe}
\frac{\partial f}{\partial t} + \mathbf{v}\bullet\nabla f - \nabla \Phi \bullet\frac{\partial f}{\partial \mathbf{v}} = 0.
\end{equation}
\noindent
Also known as the Vlasov formula, Equation (\ref{eqn:cbe}) describes
the detailed conservation of the phase-space density as an
incompressible flow under the influence of a smoothly varying
potential $\Phi(x,t)$ and represents the 1-particle limit of
Liouville's theorem.  The underlying phase-space structure of a
dissipationless system is then a conserved, fundamental quantity.

\begin{figure}
\figurenum{12}
\epsscale{1.2}
\plotone{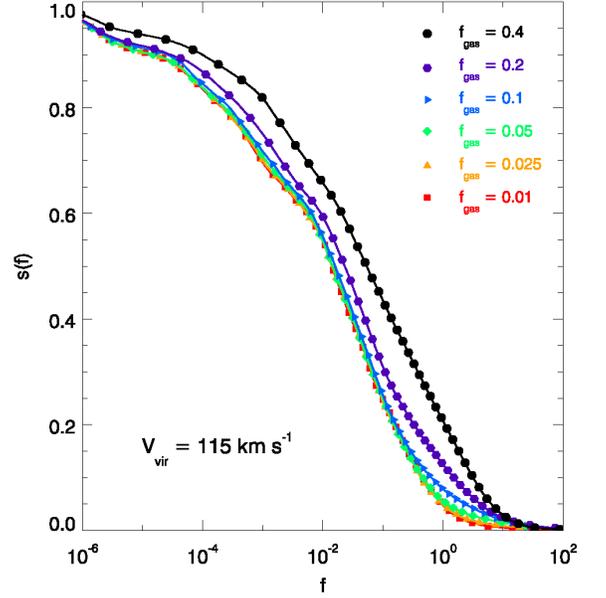}
\caption{\label{fig:df.vc2}
\small
Cumulative coarse-grained distribution function
$\ccdf$ of a remnant from the equal-mass merger of progenitors with 
$\Vvir=115$ km s$^{-1}$ as a function of progenitor gas fraction.  Shown 
is
the fraction of remnant stellar mass with phase-space densities greater 
than $\cdf$ for mergers with $\fgas = 0.01$ (red line), $\fgas = 0.025$ 
(orange line), $\fgas = 0.05$ (green line), $\fgas = 0.1$ (blue line), 
$\fgas = 0.2$ (purple line), and $\fgas = 0.4$ (black line).  For gas 
fractions $\fgas<0.2$, the phase-space density distributions of the 
remnants do not change appreciably; the binding energies of the stars do
not increase greatly relative to those in a collisionless merger as 
dissipation is not efficient at altering the structure of the galaxy.  For
gas fraction $\fgas>0.2$, the phase-space density distribution of the
remnant shifts toward higher values and more strongly-bound energy
hypersurfaces.  For low mass ellipticals the phase-space density of
remnants is then increased in gas-rich $\fgas > 30\%$ mergers where 
dissipation is important for the remnant properties. 
}
\end{figure}

\begin{figure}
\figurenum{13}
\epsscale{1.2}
\plotone{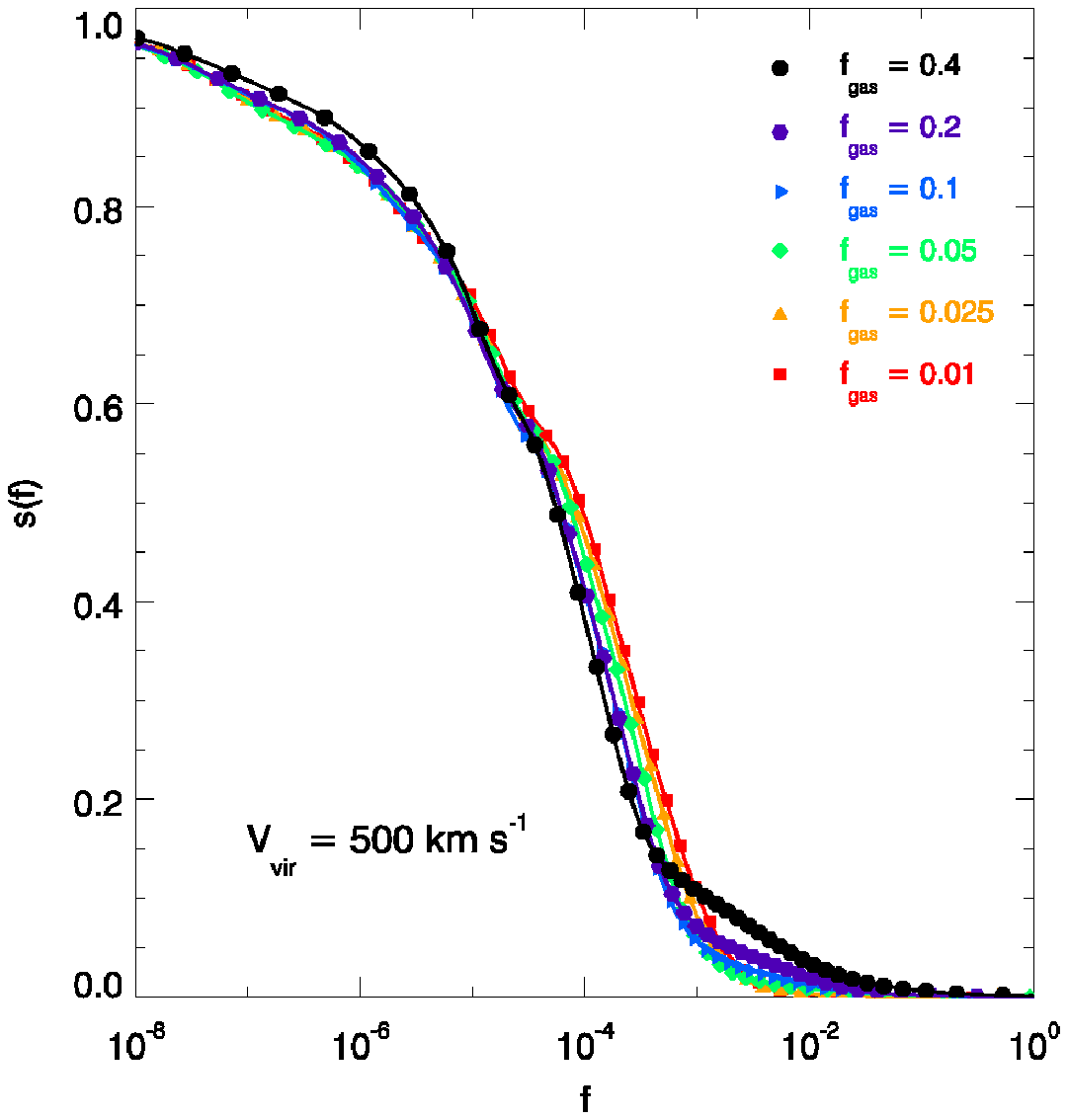}
\caption{\label{fig:df.vc6}
\small
Cumulative coarse-grained distribution function
$\ccdf$ of a remnant from the equal-mass merger of progenitors with 
$\Vvir=500$ km s$^{-1}$ as a function of progenitor gas fraction.  Shown 
is
the fraction $\ccdf$ of remnant stellar mass with phase-space densities
greater 
than $\cdf$ for mergers with $\fgas = 0.01$ (red line), $\fgas = 0.025$ 
(orange line), $\fgas = 0.05$ (green line), $\fgas = 0.1$ (blue line), 
$\fgas = 0.2$ (purple line), and $\fgas = 0.4$ (black line).  For gas 
fractions $\fgas<0.2$, the phase-space density distributions of the 
remnants do not change appreciably; the binding energies of the stars do
not increase greatly relative to those in a collisionless merger as 
dissipation is not efficient at altering the structure of the galaxy.  For
a gas fraction $\fgas>0.2$ the phase-space density distribution of the
remnant increases at moderate phase-space densities relative to the 
dissipationless remnant, but not dramatically.  The comparatively small
shift in $\ccdf$ for massive ellipticals formed in gas-rich mergers 
relative to the larger resultant shift in low mass ellipticals results in
a mass-dependent trend in $\fmtl$ (see Figures \ref{fig:mtl.vc2}--
\ref{fig:mtl.vc6}).
}
\end{figure}

Although the fine-grained phase-space density of stars is a basic
physical property of a galaxy, the finite numerical resolution of
simulations limits the measurable phase-space density to the quantity 
$\cdf(x,v)\Delta x \Delta v$ that represents the probability of a
macroscopic stellar mass element $\Delta \Mstar$ being located within the
macroscopic phase-space volume $\Delta x \Delta v$ centered at $(x,v)$.
The quantity $\cdf$ is commonly referred to as the coarse-grained 
distribution function \cite[e.g.][]{binney1987a}.
While $\cdf(x,v)$ can be measured from the simulations by gridding the
stellar particles
into $\Delta x \Delta v$ volumes, a well-sampled $\cdf(x,v)$
requires a crude $\Delta x \Delta v$ binning.
Instead, one may adopt the approach of \cite{hernquist1993b} and measure
the coarse-grained phase-space density $\cdf(E)$ averaged over 
hypersurfaces with energy 
$E$ under the assumptions of sphericity and isotropy.  
With these approximations, the fine grained phase-space density $f$ is a 
function of energy and given a model of a 
galaxy that
provides the density of states $g(E)$, the function describing
the phase-space volume of energy hypersurfaces, one may relate the 
mass-energy 
distribution of stars $\mathrm{d}\Mstar/\mathrm{d}E$ to the 
fine-grained phase-space density $f(E)$ as
\begin{equation}
\label{eqn:energy_histogram}
\frac{\mathrm{d}\Mstar}{\mathrm{d}E} = f(E)g(E).
\end{equation}
\noindent
A coarse grained distribution function $\cdf(E)$ can be measured from the
simulations through a histogram of stellar particle energies 
$\Delta \Mstar/\Delta E$ as 
\begin{equation}
\label{eqn:coarse_psd}
\cdf(E) =\frac{1}{g(E)}\frac{\Delta\Mstar}{\Delta E}.
\end{equation}
\noindent
As \cite{hernquist1993b} demonstrate, the coarse-grained $\cdf(E)$ 
provides a
reliable estimate of the fine-grained $f(E)$ for numerical models of
isotropic, spherical galaxies.  An even more precisely measurable  
quantity is the cumulative distribution function
\begin{equation}
\label{eqn:cdf}
s(f) = 
\frac{1}{\Mstar}
\int_{E_{0}}^{E(f)} \frac{\mathrm{d}\Mstar}{\mathrm{d}E}\mathrm{d}E ,
\end{equation}

\noindent
where $E_{0}$ is the central binding energy.  The cumulative distribution
function $s(f)$ provides the fraction of stellar mass at phase-space
densities above $f$, which is equivalent to the fraction of stars more
tightly bound than energy $E(f)$ when averaged over energy hypersurfaces. 
For a binned energy histogram $\Delta \Mstar/\Delta E$ measured from the
simulations, a cumulative coarse grained distribution function $s(\cdf)$ 
can be calculated as
\begin{equation}
\label{eqn:ccdf}
s(\cdf) = \frac{1}{\Mstar}\sum_{i = j_{E_{0}}}^{ j_{E(\cdf)}} \frac{\left(\Delta \Mstar\right)_{i}}{\Delta E} \Delta E ,
\end{equation}
where $\left(\Delta \Mstar\right)_{i}$ is the stellar mass with energy
between $E_{i}$ and $E_{i}+\Delta E$, $j_{E_{0}}=0$ is the index of
the energy bin with $E=E_{0}$ and $j_{E(\cdf)}$ is the index of the
energy bin with energy $E = E(\cdf)$.  The cumulative energy
distribution $s(E)$ can be measured directly from the simulations,
independent of any assumed model for $g(E)$.  The choice of galaxy
model then simply correlates $\cdf$ with an energy $E$.

\begin{figure}
\figurenum{14}
\epsscale{1.2}
\plotone{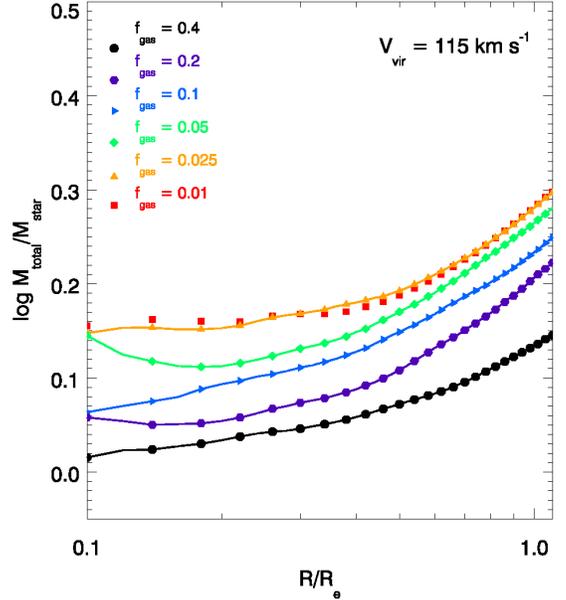}
\caption{\label{fig:mtl.vc2}
\small
Ratio of total mass $\Mtotal$ to stellar mass $\Mstar$ within a spherical 
radius for the remnants of equal mass mergers of $\Vvir = 115$ km s$^{-1}$
disk galaxies with dark matter halos, star formation and supernova 
feedback as a function of progenitor gas content.
The galaxy models are appropriate for $z=0$ and are merged on nearly 
radial, parabolic orbits.
Shown is $\fmtl$ vs. $R/\re$ for remnants produced by progenitor systems 
with gas fraction $\fgas = 0.01$ (red line), $\fgas = 0.025$ (yellow 
line), $\fgas = 0.04$ (green line), $\fgas = 0.1$ (blue line), 
$\fgas = 0.2$ (purple line), and $\fgas = 0.4$ (black line).
For small-mass spheroidal remnants, as the progenitor gas content is 
increased the stellar content substantially increases in the very central 
regions ($R < 0.3\re$) of the galaxy.
The decrease in $\fmtl$ follows the corresponding decrease in $\re$ as 
star formation is concentrated in the high central gas overdensities 
induced by the merger.
}
\end{figure}

\begin{figure}
\figurenum{15}
\epsscale{1.2}
\plotone{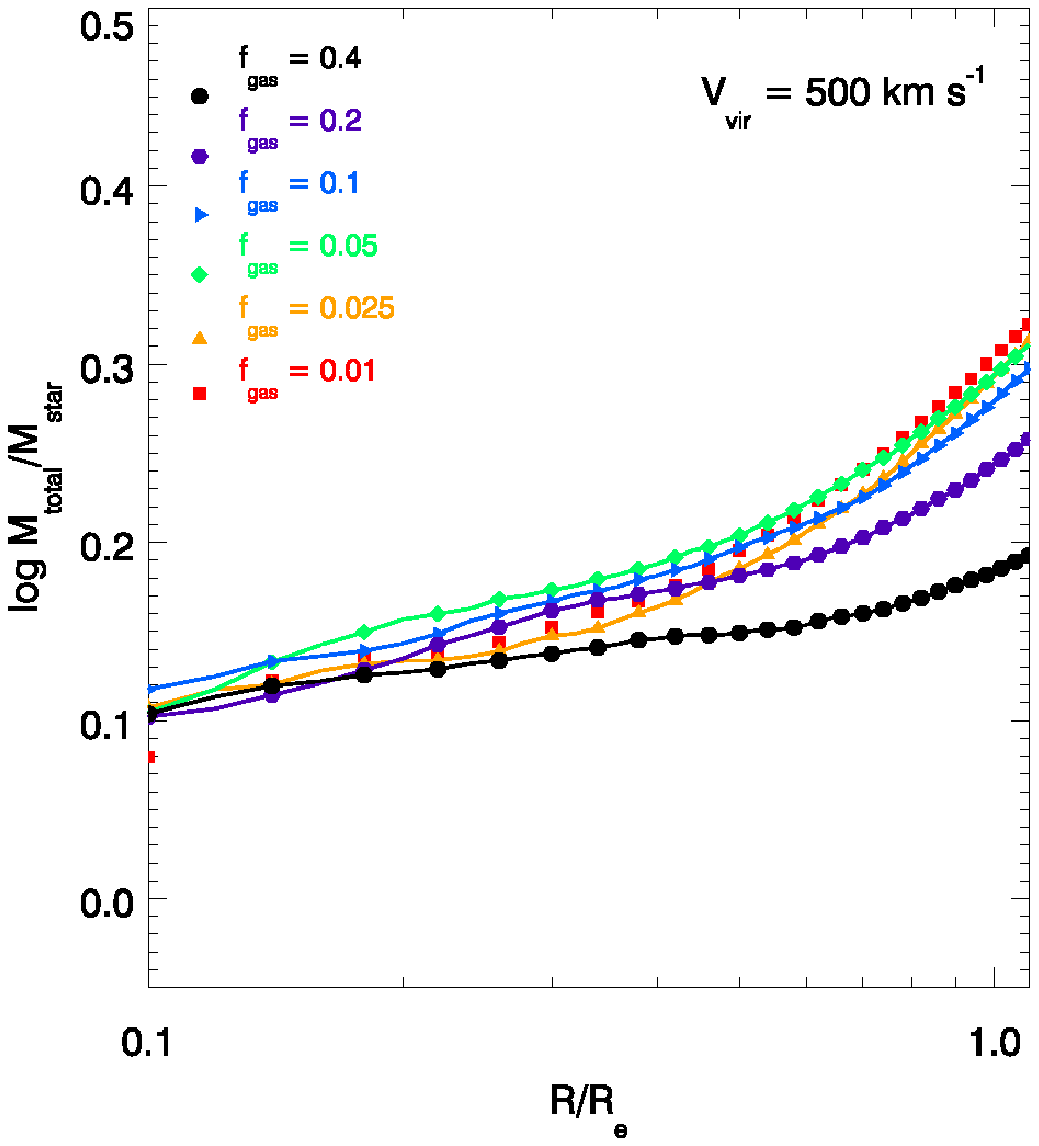}
\caption{\label{fig:mtl.vc6}
\small
Ratio of total mass $\Mtotal$ to stellar mass $\Mstar$ with spherical 
radius for the remnants of equal mass mergers of $\Vvir = 500$ km s$^{-1}$
disk galaxies with dark matter halos, star formation and supernova 
feedback as a function of progenitor gas content.
The galaxy models are appropriate for $z=0$ and are merged on nearly 
radial, parabolic orbits.
Shown is $\fmtl$ vs. $R/\re$ for remnants produced by progenitor systems 
with gas fraction $\fgas = 0.01$ (red line), $\fgas = 0.025$ (yellow 
line), $\fgas = 0.04$ (green line), $\fgas = 0.1$ (blue line), 
$\fgas = 0.2$ (purple line), and $\fgas = 0.4$ (black line).
For large-mass spheroidal remnants, as the progenitor gas content is 
increased the central stellar content begins to increase substantially 
after $R \gtrsim 0.3\re$.
The $\fmtl$ value near the center is larger for massive systems than for 
less-massive spheroids (see Figure \ref{fig:mtl.vc2}), reflecting the 
increasing importance of dark matter at the center of massive spheroidal 
galaxies.
}
\end{figure}

To connect a coarse-grained phase-space density $\cdf$ with an energy
$E$, we adopt an isotropic, stellar \cite{hernquist1990a} spheroid
embedded within a larger isotropic, dark matter \cite{hernquist1990a}
spheroid (hereafter the HH model).  \cite{ciotti1996b} analytically
calculated both the distribution function $f(E)$ and density of states
$g(E)$ for such a system, which serves as a convenient galaxy model
for the remnants and provides a conversion between binding energy and
phase-space density for a set of remnant scale radii $a_{\star}$ and
$a_{\DM}$ and stellar and dark matter masses.  The scale radii of the
remnants can be related to the effective radius and reliably estimated
from the particle distributions (e.g. \S
\ref{subsection:methodology:analysis}), while the mass of the stellar
and dark matter components are tracked explicitly.  However, as the
\cite{hernquist1990a} model may not perfectly reflect the stellar
distribution, the potential well depth at the center of the merger
remnant often exceeds the expected potential minimum for the HH model.
In this case we find the effective $a_{\star}$ for the HH system
that produces the correct minimum potential and use this model 
to calculate $f(E)$
and $g(E)$.  As the functional form of $f(E)$ for the HH model depends
only on the ratios $\Mstar/\Mdm$ and $a_{\star}/a_{\DM}$ this
technique for measuring $s(\cdf)$ is fairly insensitive to the manner
in which we measure galaxy properties.  Once violent relaxation
reduces the time-dependence of the galaxy potential during a
dissipationless merger, the form of $\ccdf$ will be fixed, while for
gaseous mergers, the form of $\ccdf$ will stabilize after the
additional process of star formation becomes relatively unimportant.
A comparison of the $\ccdf$ resulting from strongly dissipative
mergers with the $\ccdf$ of dissipationless remnants will indicate the
impact of dissipation on the final galaxy phase-space structure.

Figure \ref{fig:df.vc2} shows the cumulative coarse-grained
distribution function $\ccdf$ of a remnant from the equal-mass merger
of progenitors with $\Vvir=115$ km s$^{-1}$ as a function of
progenitor gas fraction.  The measured $\ccdf$ for gas fractions below
$\fgas<0.2$ remains roughly constant.  The binding energies of the
stars are not strongly affected by the weak influence of dissipation
on the galaxy structure for low gas fraction merger, but as the
progenitor gas fractions increase beyond $\fgas>0.2$ the phase-space
density distribution of the remnant shifts toward higher values and
more strongly-bound energy hypersurfaces.  Strongly dissipational
mergers clearly lead to higher phase-space densities in low mass
elliptical galaxy remnants.  We note here that the shapes of the
cumulative phase-space density as a function of mass and gas fraction
remain very similar for either ISM pressurization we study in this
work ($\qEOS=0.25$ and $\qEOS=1.0$), except for a slight decrease in
the maximum phase-space density reached in the $\qEOS=1.0$
simulations.

For the impact of dissipation to affect the FP tilt or $\remstar$
slope, the degree to which the phase-space density increases during
gas-rich mergers must vary as a function of galaxy mass.  Figure
\ref{fig:df.vc6} shows the cumulative coarse grained distribution
function $\ccdf$ for remnants of a fiducial equal-mass merger between disk
galaxies with $\Vvir=500$ km s$^{-1}$ as a function of progenitor gas
fraction.  While the behavior for low gas fractions $\fgas<0.2$ is
similar between low mass ellipticals (Figure \ref{fig:df.vc2}) and
massive ellipticals, the strong shift seen in the phase-space density
distribution of low mass ellipticals is not measured for the more
massive system.  The simulations show that the impact of dissipative
physics on the phase-space density of ellipticals is mass-dependent,
being more important for low mass galaxies.

\begin{figure}
\figurenum{16}
\epsscale{1.2}
\plotone{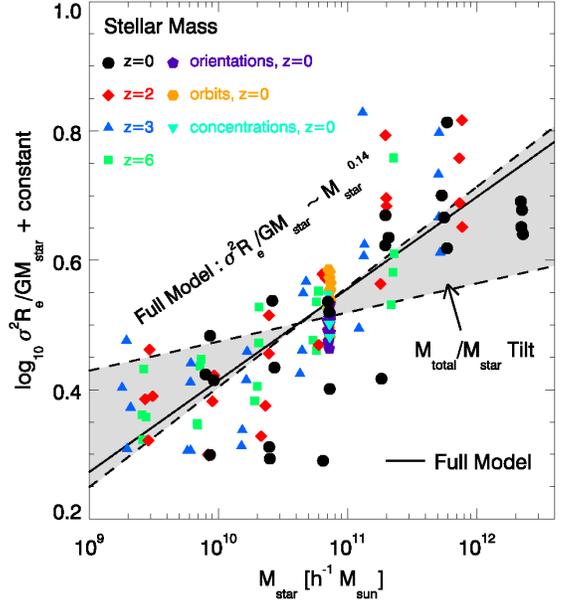}
\caption{\label{fig:mdyn}
\small
Ratio of the dynamical mass estimator $\Mdyn \equiv \sigma^{2}\re/G$ to the stellar
mass $\Mstar$ in the full-model simulations as a function of stellar mass for remnants of mergers
between progenitor disk galaxies appropriate for redshifts $z=0$ (black
circles), $z=2$ (red diamonds), $z=3$ (blue triangles), and $z=6$ 
(green squares).  Also shown is the measured $\Mdyn/\Mstar$ for $z=0$ mergers with
varying disk orientations (purple pentagons), orbital parameters (orange hexagons),
and dark matter concentrations (cyan triangles).  The best-fit trend 
$\Mdyn/\Mstar \propto \Mstar^{0.14}$ (solid line) recovers the complete fundamental plane tilt
measured for the full-model simulations in Figure \ref{fig:fp}.  We also estimate
the tilt contributed by variations of $\fmtl$ averaged within an effective radius
as a function of stellar mass measured from the remnant particle distribution (shaded
area, see \S \ref{subsubsection:results:full_model:mtl}), which contributes
$\approx40-100\%$ of the nonhomology-related FP tilt. The $\fmtl$ tilt 
estimates shown have been renormalized by a constant to intersect with the 
best-fit $\Mdyn/\Mstar$ relation at the same location.
}
\end{figure}

\subsection{Central Mass-to-Light Ratio}
\label{subsection:gas_fraction:mtl}

With a measured mass-dependent change in phase-space density of
remnants produced in gas-rich mergers, the phase-space density can be
directly connected to changes in the FP tilt and $\remstar$ slope
induced by strongly dissipational merging.  The correspondence between
phase-space density $\cdf$ and binding energy demonstrates that the
regions at the center of the galactic potential are gaining stellar
mass density.  The gas dissipation has a less dramatic effect on the
dark matter distribution of the galaxy, and while the gas does change
the dark matter profile by gravitationally dragging the dark matter
inwards, the effective radius of the dark matter decreases by only $250
h^{-1}$ pc ($0.6\%$) between the $\fgas=0.1$ and $\fgas=0.4$ runs for $\Vvir =
115$ km s$^{-1}$ halos.  Dissipation affects the stellar mass much
more dramatically than the dark matter mass, and the net result is a
radius-dependent change in the $\fmtl$ ratio as a function of
progenitor gas fraction.

Figure \ref{fig:mtl.vc2} shows the ratio of total mass $\Mtotal$ to
stellar mass $\Mstar$ within a spherical radius $R$ for the binary
$\Vvir = 115$ km s$^{-1}$ mergers with various progenitor gas fractions,
$\fgas$.  As the effects of dissipation increase, the central regions
of the low-mass remnant become progressively more baryon-dominated.
The decrease in the central $\fmtl$ follows the corresponding decrease
in $\re$ as star formation is concentrated in the high central gas
overdensities induced by the merger.  The value of $\fmtl$ does not
change appreciably at $R=\re$ until the gas fraction reaches $\fgas =
0.1$ and decreases dramatically as the gas fraction increases toward
$\fgas=0.4$.

The impact of dissipation on $\fmtl$ of more massive systems is less
important, as expected from the behavior of the phase-space density
and $\re$ of those systems.  Figure \ref{fig:mtl.vc6} shows the ratio
$\fmtl$ with spherical radius $R$ for the remnants of equal mass
mergers of $\Vvir = 500$ km s$^-1$ systems with various gas fractions.
For high-mass spheroidal remnants, as the progenitor gas content is
increased, the stellar content begins to increase substantially at
radii $R > 0.3\re$.  The central $\fmtl$ remains relatively unaffected
by dissipational physics and remains roughly constant
$\fmtl\approx1.25$.  The value of $\fmtl$ near the center is larger
for massive systems than for less-massive spheroids (see Figure
\ref{fig:mtl.vc2}), reflecting the increasing importance of dark
matter at the center of massive spheroidal galaxies and contributing
tilt to the FP.

\subsubsection{Dynamical Mass-to-Light Ratio}
\label{subsubsection:results:full_model:mtl}
For the stellar-mass FP relation, an observationally relevant quantity is
the estimator of the dynamical mass within an $\re$
\begin{equation}
\label{eqn:m_dyn}
\Mdyn \equiv k\frac{\sigma^{2}\re}{G}
\end{equation}
\noindent
where $G$ is the gravitational constant and $k$ is a constant 
that depends on the structure of the galaxy ($k\approx7.25$ for a 
kinematically isotropic \cite{hernquist1990a} spheroid).  The dynamical
mass estimator allows us to characterize the FP tilt in terms of the 
observationally-accessible mass ratio
\begin{equation}
\label{eqn:mtl_dyn}
\frac{\Mdyn}{\Mstar} \propto \Mstar^{\gamma}.
\end{equation}
\noindent
For a stellar-mass FP of the form $\re \propto \sigma^{\alpha} \Ie^{-\beta}$ 
and an
$\remstar$ relation of the form $\re \propto \Mstar^{\mu}$, the 
power-law index $\gamma$ can be estimated as
\begin{equation}
\gamma = \mu(2/\alpha+1-4\beta/\alpha)+2\beta/\alpha-1.
\end{equation}
For the virial scalings ($\alpha=2$, $\beta=1$), the dynamical mass
estimator is proportional to the stellar mass and there is no FP
tilt ($\gamma=0$), independent of the $\remstar$ relation.  The results
presented in \S \ref{subsection:results:full_model} indicate that 
the full-model FP ($\alpha=1.55$, $\beta=0.82$)
is tilted relative to the virial scalings.  Combined with the 
full-model $\remstar$ relation ($\mu=0.57$), we expect that
$\Mdyn/\Mstar \propto \Mstar^{0.15}$.  Figure \ref{fig:mdyn} shows
the quantity $\Mdyn/\Mstar$ as a function of stellar mass for the
full-model remnants from mergers of progenitor disk galaxies appropriate
for redshifts $z=0-6$.  While there is considerable scatter in 
$\Mdyn/\Mstar$, which mostly arises from the linear dependence of
$\Mdyn$ on $\re$, the best 
fit trend in the simulations has a scaling $\Mdyn/\Mstar\propto\Mstar^{0.14}$
(solid line) that recovers essentially the complete tilt inferred for the
full-model FP.

In principle, the tilt in the stellar-mass FP inferred from the 
dynamical mass estimator may 
include contributions from total mass-to-stellar mass $\Mtotal/\Mstar$
variations, kinematic
nonhomologies that affect the velocity dispersion as a function of 
stellar mass, or other sources of nonhomology.
Figure \ref{fig:mdyn} shows the estimated
range of tilt contributed by $\fmtl$ variations with stellar mass 
averaged over an effective radius as measured from the particle 
distributions of the remnants (shaded area), bounded by least-squares
fits to the $\fmtl$ trend with mass minimizing with respect to
either $\fmtl$ or $\Mstar$.  Based on this comparison, we estimate
that variations of $\fmtl$ 
with stellar mass averaged over an effective radius contribute
$\approx40-100\%$ of the nonhomology-related tilt in the 
full-model simulations.

\begin{figure}
\figurenum{17}
\epsscale{1.2}
\plotone{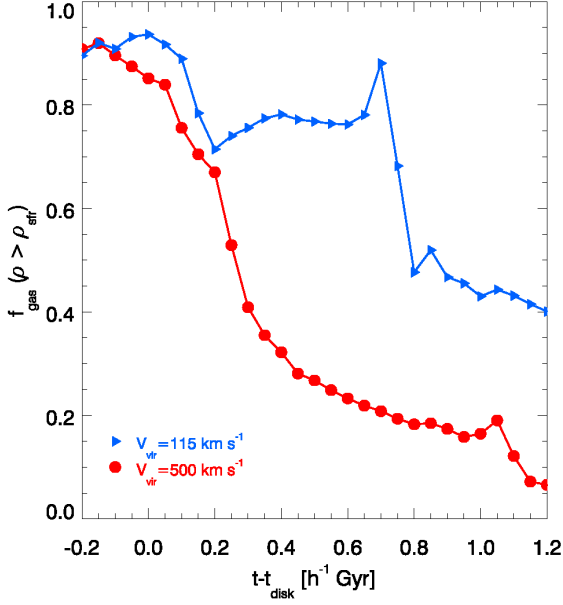}
\caption{\label{fig:gf.sfr}
\small
Mass fraction of star forming gas for $\Vvir=115$ km s$^{-1}$
(blue triangles) and $\Vvir=500$ km s$^{-1}$ (red circles) progenitor
merger remnants in simulations with gas fraction $\fgas=0.4$ as a function
of time $t-\tdisk$ since the disks first interact.  The gas in the massive
system is collisionally heated during the first passage and subsequently
cannot cool and form stars, while gas in the small mass system can cool
sufficiently to continue to form stars throughout the remainder of the 
merger.
}
\end{figure}

\begin{figure}
\figurenum{18}
\epsscale{1.2}
\plotone{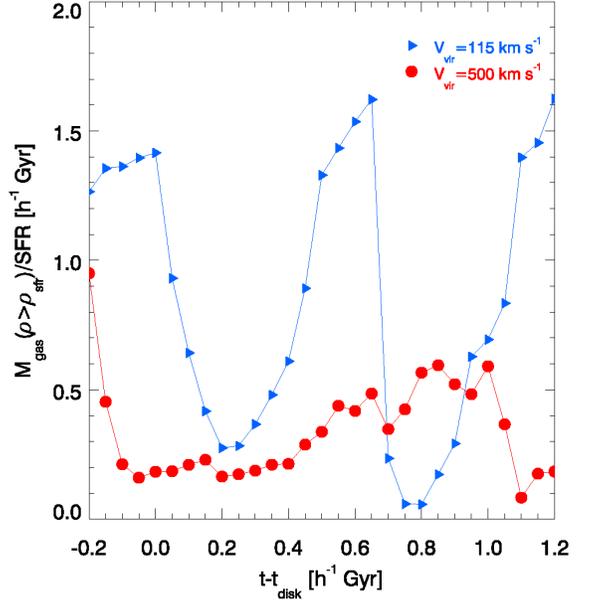}
\caption{\label{fig:gf.consumption}
\small
Consumption timescale of star forming gas in $\Vvir=115$ km 
s$^{-1}$ (blue triangles) and $\Vvir=500$ km s$^{-1}$ (red circles) 
progenitor mergers measured for simulations with gas fraction $\fgas=0.4$ 
and equation of state parameter $\qEOS=0.25$.  The consumption timescale
is plotted as a function of time $t-\tdisk$ since the disks first 
interact and reflects the ratio of 
$\Mgas(\rho>\rho_{\mathrm{sfr}})/(\mathrm{d}\Mstar/\mathrm{d}t)$ during 
the simulation for gas above the density threshold for
star formation.  The cold
gas in the massive system that is not collisionally heated by the first
pericentric passage into the halo (see Figure \ref{fig:gf.sfr}) is 
always efficiently converted into stars.  In smaller systems, the cold
gas in the system is efficiently converted into stars only during times
of interactions, such as pericentric passage ($t-\tdisk = 0.1-0.2$ 
h$^{-1}$ Gyr) or final coalescence ($t-\tdisk\approx0.9$ h$^{-1}$ Gyr), 
but is 
consumed on long timescales otherwise.
}
\end{figure}

\subsection{Dissipation vs. Galaxy Mass}
\label{subsection:gas_fraction:dissipation}

The previous sections have demonstrated that gas dissipation impacts
the fundamental scaling properties of ellipticals by altering their
central stellar content in a mass-dependent manner, being more
important for small mass systems.  We now explain how this
mass-dependent impact of dissipation arises in our simulations.

Figure \ref{fig:gf.sfr} shows the mass fraction of dense, star forming
gas in merging disk galaxies as a function of the time $t-\tdisk$
since the gaseous disks first interact for a simulation with
progenitor gas fraction $\fgas=0.4$ and equation of state parameter
$\qEOS = 0.25$.  In massive systems, such as the $\Vvir=500$ km
s$^{-1}$ progenitor equal-mass merger shown here (red circles), a
large fraction of the gas is collisionally heated during the first
pericentric passage.  This hot gas becomes x-ray luminous but cannot
cool efficiently \citep{cox2004a,cox2005a}, and mostly remains in the
halo for the duration of the merger.  The efficiency of halo gas
cooling decreases with halo mass, causing more massive systems to act
increasingly like dissipationless systems.  In small systems (we show
an equal-mass merger of $\Vvir=115$ km s$^{-1}$ progenitors, blue
triangles), collisionally heated gas can cool efficiently and return
to the disk to continuously form stars.  The continued presence of
star forming gas in small mass systems allows for very high
phase-space density stars to form in abundance during the height of
the merger, decreasing the central $\fmtl$ in the final remnant.

Although collisional heating of the gas in massive systems limits the
fraction of gas dense enough to form stars, the remaining dense gas is
efficiently turned into stars throughout the merger.  Figure
\ref{fig:gf.consumption} shows the consumption timescale
\begin{equation}
t_{\mathrm{consumption}} = \Mgas(\rho>\rho_{\mathrm{sfr}})\left(\frac{\mathrm{d}\Mstar}{\mathrm{d}t}\right)^{-1}
\end{equation}
\noindent
which indicates the characteristic timescale for the mass of gas at
densities above the star formation threshold
$\Mgas(\rho>\rho_{\mathrm{sfr}})$ to be converted into stars at the
current star formation rate $\mathrm{d}\Mstar/\mathrm{d}t$.  The
massive system is always efficiently converting the available gas into
stars, roughly independent of the progenitor separation after
first approach when tidal deformation of the disk begin vigorous star
formation.  We note that while the star formation is always efficient,
the massive system, whose original progenitors had gas fractions of
$\fgas = 0.4$, becomes largely dissipationless after $t-\tdisk = 0.4$
$h^{-1}$ Gyr when the total gas fraction drops below $\fgas=0.06$
and the amount of cold, star forming gas is a fraction $\sim0.30$ 
of the gas mass.
Dissimilarly, the small mass system experiences large variations in
its consumption timescale, with extremely efficient star formation at
the first pericentric passage ($t-\tdisk\approx0.1-0.2$ $h^{-1}$ Gyr)
and the final coalescence ($t-\tdisk\approx0.85$ $h^{-1}$ Gyr) and
inefficient star formation at larger progenitor separations and after
the merger is completed.  We comment that this variation in
consumption timescale for small mass systems differs importantly from
the scenario described by \cite{bekki1998a}, who infers that tilt in
the fundamental plane may arise from a gradation in consumption
timescale or star formation ``rapidity'' with galaxy mass.  While the
time-averaged consumption timescale in smaller systems is longer as
emphasized by \cite{bekki1998a}, we do not introduce an additional ad
hoc mass-dependency to the gas consumption timescale.  Differences in
the gas consumption timescale as a function of galaxy mass in our
modeling owe to the increase in disk gas density with galaxy mass,
with low mass systems naturally having comparatively inefficient star
formation relative to massive systems \emph{in quiescence}.  However,
during phases of strong interaction, the gas density in low mass
systems increases dramatically and leads to a rapid decrease in the
gas consumption timescale.  The dynamism of the consumption
timescale in small mass systems then allows the less massive system to
``save'' its gas for strong interactions, when dissipation is very
effective at converting gas into stars in the central-most regions on
a short timescale \citep[e.g.][]{mihos1994a}.

\section{Discussion}
\label{section:discussion}

Our simulations indicate that mergers of gas-rich disk galaxies
produce tilt in the Fundamental Plane relationship relative to the
scalings predicted by the virial theorem for homologous systems.  The
remnants of gas-rich mergers are not homologous as gas dissipation is
more effective at redistributing stellar material relative to dark
matter in smaller systems than in larger ones.  Strongly dissipational
mergers also appear necessary to steepen the $\remstar$ relation of
disk galaxy remnants to more closely follow the observed
\cite{shen2003a} relation for early-type galaxies.

Most conservatively interpreted, our results demonstrate simply that
if elliptical galaxies originate in gas-rich mergers then they will
obey a tight fundamental plane relation that has significant tilt
owing to dissipative effects including central $\fmtl$ variations 
as a function of galaxy mass.
However, independent observational and theoretical evidence suggests
that gas-rich mergers are a necessary feature of the merger hypothesis
as an elliptical galaxy formation scenario.

\cite{bender1992a} provided an early characterization of the
importance of dissipation varying with the mass of elliptical
galaxies. In their ``gas/stellar'' (GS) continuum, massive systems
experience less dissipation, slowly rotate, and are increasingly
flattened by anisotropy.  Our work describes the impact of the GS
continuum on the central structure of elliptical galaxies as a
function of their stellar mass and details the origin of the GS
continuum in terms of the inefficient cooling of collisionally heated
gas in massive systems and dynamic variations in the gas consumption
timescale in smaller mass mergers.  As we note in \S
\ref{subsection:gas_fraction:dissipation}, these effects differ
substantially in detail from previous suggestions that the gas
consumption timescale was important for elliptical FP properties
\citep[e.g.][]{bekki1998a}.  Other theoretical studies of elliptical
galaxy formation have suggested that gas dissipation was necessary for
setting or maintaining elliptical galaxy scaling laws, notably
\cite{ciotti2001a} who conclude from dissipationless mergers that
significant dissipation is required to simultaneously satisfy both the
$\msigma$ and FP relation.  Combined with the results of
\cite{di_matteo2005a} and \cite{robertson2005a}, this work
demonstrates that highly dissipational mergers $can$ generate both the
FP and $\msigma$ relations with tight scatter.  Our analysis has also
provided a detailed explanation of how dissipation systematically
varies with galaxy mass and to what extent it induces FP tilt.

Furthermore, our dissipationless simulations are consistent
with the emerging picture that at least one generation of spheroidal
mergers do not strongly invalidate the scaling laws of their
elliptical galaxy progenitors
\citep{capelato1995a,dantas2003a,nipoti2003a,boylan-kolchin2005a}.
The recent observations suggesting that elliptical galaxies undergo
$0.5-1$ such mergers at redshifts $z<1$
\citep{bell2005a,van_dokkum2005a} may require such behavior as the
tight local FP relation observationally limits the possible importance
of scenarios that substantially increase the FP scatter.  We also note
here that the results from our simulations of the $\msigma$ relation
produced by re-mergers of ellipticals are in rough agreement with
previous studies \citep[e.g.][]{nipoti2003a}.  Moreover, the fact that 
re-merger remnants do not increase in velocity dispersion dramatically
while their black hole masses double suggests that if black holes are
overly massive relative to their host bulges at high redshifts
\citep[e.g.][]{walter2004a,peng2005a}, then major dissipationless
mergers of spheroids will only exacerbate their discrepancy from the
local $\msigma$ relation
\citep{gebhardt2000a,ferrarese2000a,tremaine2002a}.  If the structural
properties of quiescent galaxies instead increase the central velocity
dispersions at a given black hole mass at high redshifts, major
dissipationless mergers of spheroids will decrease the scatter in the
$\msigma$ relation locally \citep{robertson2005b}.  Lastly, if the
frequency of major dissipationless mergers increases with galaxy mass then
the high-mass end of the $\msigma$ relation may steepen relative to the
\cite{tremaine2002a} best-fit relation and
appear as a mass-dependent scatter in the 
data \citep{robertson2005b}.

Independent evidence from simulations that dissipation is important
for the properties of elliptical galaxies is mounting.  Using
dissipational simulations of disk galaxy mergers, including a subset
of the simulations analyzed in this work, \cite{cox2005b} demonstrate
that progenitor gas fractions of order $\fgas\approx0.3$ or higher are
necessary to reproduce the kinematic structure of elliptical galaxies
as a function of their stellar mass.  Notably, the $v/\sigma$ ratio of
rotation to dispersion of remnants from $\fgas=0.4$ mergers better
reproduces the observed distribution of elliptical kinematics than do
purely dissipationless disk galaxy mergers \citep[e.g.][]{naab2003a}.
Our analysis demonstrates that the \cite{cox2005b} remnants also lie
on a tight FP relation.

While recent observational determinations of the fundamental plane in
galaxies at redshifts $z\sim1$ suggest that the $M/L$ offset of
the FP changes with redshift owing to passive evolution of the stellar
populations, there is still disagreement regarding the redshift-evolution 
of the FP tilt \citep[e.g.][]{gebhardt2003a,van_dokkum2003a,van_der_wel2005a,treu2005a,di_serego_alighieri2005a}.
The results of our
simulations of disk galaxy mergers appropriate for higher redshifts
suggest that if the $M/L$ tilt induced by stellar population effects
remain fixed as a function of galaxy mass with redshift
(i.e. evolution in the stellar populations only affects the FP
zero-point), then the observed FP tilt will remained fixed out to high
redshifts.  The implication from stellar population modeling that
elliptical galaxies likely formed early ($z>1-2$) then may not provide
a tight constraint on the merger hypothesis as a theory for elliptical
galaxy formation since gas-rich disk galaxy models appropriate for a
large range in redshift produce the same stellar-mass FP relation.  As
methods for including a cosmological accounting of elliptical galaxy
formation times and stellar population color evolution into structure
formation theory improve 
\citep[see, e.g.][for a related methodology]{hopkins2005b}, theoretical 
models will have more to say in the future on this topic.

As a final note, some recent observations have been interpreted as
evidence that the FP tilt is consistent with a purely
stellar-population induced $M/L$ variation with galaxy mass
\citep[e.g.][]{cappellari2005a}.  The success of \cite{cox2005b} at
reproducing the kinematic properties of elliptical galaxies with
gas-rich disk galaxy merger simulations (including some simulations
analyzed in this paper) argues strongly that dissipation is important
for understanding kinematical mass trends.  Our simulations suggest
that a corollary of the dissipational origin of kinematic structure in
ellipticals is FP tilt, which at least partially owes to central
$\fmtl$ variations with galaxy mass.  We note that this tilt may be in
addition to both kinematic nonhomology-related and photometric
nonhomology-related tilt, and a full characterization of these effects
is planned for future work.

\section{Summary}
\label{section:summary}

We determine the Fundamental Plane and $\remstar$ relations produced
by the merging of galaxies appropriate for redshifts $z=0-6$.  We
demonstrate that gas dissipation induces tilt into the Fundamental
Plane, causing a deviation from the theoretical scalings derived from
the virial theorem for homologous systems.  Furthermore, the recovery
of the $\remstar$ relation for early-type galaxies similarly requires
gas-rich progenitors to steepen the relation relative to the
$\remstar$ relation produced by the dissipationless merging of disk
galaxies.  These simulations argue that the merger hypothesis for the
formation of locally-observed elliptical galaxies is successful if
disk galaxies merge when dissipational effects are sufficiently
important, requiring gas fractions $\fgas > 0.3$ in disk galaxies at
redshifts where ellipticals are primarily formed (typically $z>1$). We
provide a detailed summary of these results below.

\begin{itemize}

\item
The Fundamental Plane (FP) relation produced by the merging of 
dissipationless 
disk galaxy models lies nearly parallel to the plane
defined by the virial relation for homologous systems.
Increasing the angular momentum of the orbit by lengthening the 
pericentric passage distance of the orbit produces an offset in the FP by 
increasing the effective radius of the remnants.  Scaling the disk galaxy
models for redshifts $z=0-6$ produces the same FP relation.
The effective radius $\re$ -- stellar mass $\Mstar$ relation produced
in these mergers is shallower than that measured for massive galaxies
in the Sloan Digital Sky Survey \citep{shen2003a}.  Wide orbits and
progenitors appropriate for low-redshifts produce larger remnants than
do high- redshift progenitors or progenitors with bulges, but all
dissipationless mergers produce shallow $\remstar$ relations.  We also
use higher resolution simulations to demonstrate that these results
are insensitive to our numerical resolution.

\item
The merging of gas-rich disk galaxy models with cooling, star
formation, and supernova feedback produces a Fundamental Plane (FP)
scaling ($\re \propto \sigma^{1.58} \Ie^{-0.80}$) similar to the observed 
infrared FP \citep[$\re \propto \sigma^{1.53} \Ie^{-0.79}$;][]{pahre1998a}
and is
almost independent of the redshift scalings of the progenitor systems.
The effective radius $\re$ -- stellar mass $\Mstar$ relation produced
by these mergers is roughly parallel to that measured for massive
galaxies in the Sloan Digital Sky Survey \citep{shen2003a}, but 
the chosen prograde-prograde coplanar orbit generates remnants that 
lie somewhat below this relation.
Higher-redshift progenitors produce smaller remnants.

\item
The Fundamental Plane (FP) relation produced by the merging of
gas-rich disk galaxies with cooling, star formation, supernova
feedback, and a prescription for feedback from accreting supermassive
black holes exhibits a scaling ($\re \propto \sigma^{1.55} \Ie^{-0.82}$)
similar to the observed infrared FP
\citep[$\re \propto \sigma^{1.53} \Ie^{-0.79}$;][]{pahre1998a} 
and is nearly coincident with the FP produced by
similar simulations without black holes.  The FP relation is roughly
independent of the redshift scalings of the progenitor systems and the
location of remnants within the FP is fairly insensitive to a large
variety of disk orientations and orbital configurations as changes in
the effective radius are compensated by changes in the velocity
dispersion and surface mass density.  The effective radius $\re$ --
stellar mass $\Mstar$ relation produced by the same simulations
produces a $\re - \Mstar$ relation roughly parallel to that measured
for massive galaxies in the Sloan Digital Sky Survey
\citep{shen2003a}, with an offset toward smaller remnants.
Varying
the system angular momentum through combinations of the initial disk
orientation and pericentric passage distance for a single pair of
progenitor models produces a spread in the remnant effective radius, and
a proper accounting of cosmological orbits would decrease the discrepancy
with the \cite{shen2003a} normalization.
The $\re - \Mstar$ relation shifts toward smaller remnants as the
redshift of the progenitor systems increases.

\item
We estimate that $\approx 40-100\%$ of the Fundamental Plane (FP) tilt 
produced in the
simulations owes to an increase in stellar mass at high
phase-space densities and binding energies as gas converts into stars
in the central-most regions of the remnants.  This increase in
phase-space density corresponds to a decrease in the total-to-stellar
mass ratio $\fmtl$ in the central regions of simulated
elliptical galaxies.  Furthermore, the decrease in $\fmtl$ is a
function of galaxy mass, with a stronger decrease in smaller systems,
thereby inducing a FP tilt.  We find that the gas-rich mergers
($\fgas>0.3$) required to steepen the $\remstar$ relation to become
parallel with the observed \cite{shen2003a} relation also produce
significant FP tilt.

\item
The mass-dependent impact of dissipation on the central $\fmtl$ of
galaxies is tied to scale-dependent processes in the merging systems.
In massive halos, a substantial fraction of the available gas is
collisionally heated during the first pericentric passage and driven
into the halo.  This heated gas cannot cool efficiently, thereby
causing massive systems to act more like dissipationless systems than
low mass merging galaxies. Moreover, the dynamic variation of the
consumption timescale in low mass galaxies allows for the rapid
conversion of gas into stars during only the most violent interactions
of the merger while conserving the gas content of the system in more
quiescent phases and allowing for gas dissipation to have an increased
impact on the central properties of the remnants.

\item
The Fundamental Plane (FP) relation produced by the merging of
spheroidal \cite{hernquist1990a} galaxy models with dark matter halos
remains similar to the FP delineated by their progenitors, in
agreement with previous results \citep[e.g.][]{boylan-kolchin2005a}.
Additionally, the re-merging of the remnants from gas-rich disk galaxy
mergers also roughly maintains the FP of their progenitors.  The
effective radius $\re$ -- stellar mass $\Mstar$ relations produced by
the merging of spheroidal galaxy models produces a slightly shallower
$\remstar$ relation than that delineated by the progenitor models, and
although the remnants nearly double in effective radius through
equal-mass mergers, a single generation of remnant re-merging will
likely induce scatter but not destroy the $\remstar$ correlation.

\item
The re-merging of ellipticals formed from the merging of disk galaxies
will likely induce additional scatter in the local black hole mass
($\MBH$) -- stellar velocity dispersion ($\sigma$) relation, but should
not destroy the correlation after a single generation of remnant
re-merging.  In agreement with previous simulations
\citep[e.g.][]{nipoti2003a}, we find that the velocity dispersions of
the re-merger remnants do not increase substantially relative to their
progenitors during binary mergers.  However, the typical number of $0.5-1$
dissipationless mergers below $z=1$ recently inferred from
observations \citep[e.g.][]{bell2005a,van_dokkum2005a} should not
dramatically change the $\msigma$ scalings imprinted at higher
redshifts.  If the velocity dispersion in remnants for a given black
hole mass is larger at high redshifts than that observed locally, as
inferred from simulations of the generation of the $\msigma$ relation
from the merger of disk galaxies at redshifts $z=0-6$
\citep{robertson2005b}, the dissipationless re-merging of remnants
would \emph{decrease} the scatter in the local $\msigma$ relationship
under the condition that the strength of black hole feedback did not
vary significantly with redshift.  If black holes are substantially more
massive than their host bulges at high redshifts
\citep[e.g.][]{walter2004a,peng2005a}, major dissipationless mergers
of spheroids at low redshifts would \emph{exacerbate} the discrepancy
of those galaxies with the local $\msigma$ relation.  An 
increasing 
frequency of major dissipationless mergers with galaxy mass may steepen
the high-mass $\msigma$ relation relative to the local best-fit trend
\citep[e.g.][]{tremaine2002a}.

\end{itemize}

The success of the merger hypothesis for elliptical galaxy formation
seems intimately connected with the physical process of dissipation.
Our results demonstrate that if disk galaxy mergers are strongly
dissipational, tilt is generated in Fundamental Plane and the
$\remstar$ relation steepens to approach the observed scaling.
Combined with compelling evidence from the kinematic structure of
simulated elliptical galaxies \citep[e.g.][]{cox2005b}, we conclude
from our results that the formation of elliptical galaxies most likely
occurred in gas-rich mergers of disk galaxies.  Reconciling
observational determinations of the Fundamental Plane with the merger
hypothesis will require the abandonment of simplistic assumptions for
the relative stellar and dark matter mass distributions at the centers
of elliptical galaxies for the interpretation of observations and
require more sophisticated modeling of cosmological structure
formation, including a full cosmologically motivated distribution of
elliptical stellar colors and metallicities.

\acknowledgements
BR acknowledges helpful conversations with Alar Toomre, Luca Ciotti,
Joel Primack, Eliot Quataert, and Sandy Faber, and would like to thank
Mike Boylan-Kolchin for detailed comparisons of analysis techniques.
This work was supported in part by NSF grants ACI 96-19019, AST
00-71019, AST 02-06299, and AST 03-07690, and NASA ATP grants
NAG5-12140, NAG5-13292, and NAG5-13381.  The simulations were
performed at the Center for Parallel Astrophysical Computing at
Harvard-Smithsonian Center for Astrophysics.


\end{document}